\documentclass[useAMS,usenatbib,usegraphicx]{mn2e}  
\usepackage[usenames]{color}
\usepackage{url}

\newcommand{\be}{\begin{equation}}
\newcommand{\ee}{\end{equation}}
\newcommand{\bea}{\begin{eqnarray}}
\newcommand{\eea}{\end{eqnarray}}

\newcommand{\vectX}{\bf {\it X}}
\newcommand{\vectTheta}{\bf {\it \Theta}}
\newcommand{\matrGamma}{\bf \Gamma}
 \title[DM in A370] 
{A free-form lensing model of A370 revealing stellar mass dominated BCGs, in Hubble Frontier Fields images.} 

\author[J.M. Diego]  
  {Jose M. Diego\thanks{jdiego@ifca.unican.es}$^1$, 
   Kasper B. Schmidt$^{2}$, 
   Tom Broadhurst$^{3,4}$, 
   Daniel Lam$^5$,
  \newauthor
   Jes\'us Vega-Ferrero$^{6,11}$,
   Wei Zheng$^{7}$,
   Slanger Lee$^{12}$,
   Takahiro Morishita$^{8,9,10}$, 
  \newauthor
   Gary Bernstein$^{11}$,
   Jeremy Lim$^{12,13}$,
   Joseph Silk$^{7,14,15,16}$, 
   Holland Ford$^7$\\
$^{1}$ IFCA, Instituto de F\'isica de Cantabria (UC-CSIC), Av. de Los Castros s/n, 39005 Santander, Spain\\
$^{2}$ Leibniz-Institut für Astrophysik Potsdam (AIP), An der Sternwarte 16, 14482 Potsdam, Germany\\
$^3$  Fisika Teorikoa, Zientzia eta Teknologia Fakultatea, Euskal Herriko Unibertsitatea UPV/EHU, E-48080 Bilbao, Spain\\   
$^4$ IKERBASQUE, Basque Foundation for Science, Alameda Urquijo, 36-5 48008 Bilbao, Spain\\
$^5$ Leiden Observatory, Leiden University, NL-2300 RA Leiden, The Netherlands\\
$^6$ Departamento de F\'isica Te\'orica, Universidad Aut\'onoma de Madrid, 28049 Madrid, Spain\\
$^7$ Department of Physics and Astronomy, The Johns Hopkins University Homewood Campus, Baltimore, MD 21218, USA\\ 
$^{8}$ Department of Physics and Astronomy, University of California, Los Angeles, CA 90095-1547, USA\\
$^{9}$ Astronomical Institute, Tohoku University, Aramaki, Aoba, Sendai 980-8578, Japan\\
$^{10}$ Institute for International Advanced Research and Education, Tohoku University, Aramaki, Aoba, Sendai 980-8578, Japan\\
$^{11}$ Department of Physics and Astronomy, University of Pennsylvania, 209 S. 33rd St, Philadelphia, PA 19104, USA\\
$^{12}$ Department of Physics, The University of Hong Kong, 0000-0002-6536-5575, Pokfulam Road, Hong Kong\\
$^{13}$ Laboratory for Space Research, Faculty of Science, The University of Hong Kong, 0000-0002-6536-5575, Pokfulam Road, Hong Kong\\
$^{14}$ Institut d\'Astrophysique de Paris (UMR 7095: CNRS \& UPMC - Sorbonne Universités), 98 bis bd Arago, F-75014 Paris, France\\
$^{15}$ Laboratoire AIM-Paris-Saclay, CEA/DSM/IRFU, CNRS, Univ. Paris VII, F-91191 Gif-sur-Yvette, France\\
$^{16}$ BIPAC, Department of Physics, University of Oxford, Keble Road, Oxford OX1 3RH, UK
}

\date{Draft version \today}  
  
\pagerange{\pageref{firstpage}--\pageref{lastpage}}  
  
\begin{document}  
\maketitle  
 
\label{firstpage}  
%%%%%%%%%%%%%%%%%%%%%%%%%%%%%%%%%%%%%%%%%%%%%%%%%%%%%%%%%%%%%%%%%%%%%%%%%%%%%%%  
\begin{abstract} 

We derive a free-form mass distribution for the unrelaxed cluster A370 ($z=0.375$), using the latest  Hubble Frontier Fields images and GLASS spectroscopy. Starting from a reliable set of 10 multiply lensed systems we produce a free-form lens model that identifies $\approx 80$ multiple-images. Good consistency is found between models using independent subsamples of these lensed systems, with detailed agreement for the well resolved arcs. The mass distribution has two very similar concentrations centred on the two prominent Brightest Cluster Galaxies (or BCGs), with mass profiles that are accurately constrained by a uniquely useful system of long radially lensed images centred on both BCGs. We show that the lensing mass profiles of these BCGs are mainly accounted for by their stellar mass profiles, with a modest contribution from dark matter within $r<100$ kpc  of each BCG. This conclusion may favour a cooled cluster gas origin for BCGs, rather than via mergers of normal galaxies for which dark matter should dominate over stars. Growth via merging between BCGs is, however, consistent with this finding, so that stars still dominate over dark matter .

\end{abstract}  
%%%%%%%%%%%%%%%%%%%%%%%%%%%%%%%%%%%%%%%%%%%%%%%%%%%%%%%%%%%%%%%%%%%%%%%%%%%%%%%  
%%%%%%%%%%%%%%%%%%%%%%%%%%%%%%%%%%%%%%%%%%%%%%%%%%%%%%%%%%%%%%%%%%%%%%%%%%%%%%%  

%%%%%%%%%%%%%%%%%%%%%%%%%%%%%%%%%%%%%%%%%%%%%%%%%%%  
\section{Introduction}\label{sect_intro}  
%%%%%%%%%%%%%%%%%%%%%%%%%%%%%%%%%%%%%%%%%%%%%%%%%%%  

The Hubble Frontier Fields program\footnote{http://www.stsci.edu/hst/campaigns/frontier-fields/} \citep[or HFF hereafter, ][]{Lotz2016} provides the most remarkably detailed examples  of gravitational lensing by galaxy clusters, registering hundreds of multiply-lensed galaxies for charting galaxy formation to unprecedented depths \citep[see e.g ][]{Lam2014,Diego2015a,Diego2015b,Diego2016}. Furthermore, most of these HFF clusters are in a state of collision, enhancing their value for assessing the collisionality of dark matter, a basic assumption of the standard particle interpretation of dark matter  \citep{Markevitch2002,Markevitch2004,Springel2007,Randall2008}. Many clusters exhibit significant, but modest, offsets between the peak of the dark matter distribution and the centroid of  the X-ray emission  \citep{Markevitch2004,Clowe2006,Mahdavi2007,Menanteau2012,Dawson2012}, which is expected if dark matter is collisionless. These observations can provide a constraint on the dark matter cross-section \citep{Markevitch2004,Randall2008}. It is important that these differences are evaluated with the guidance of hydrodynamical models, as complex multi-body interactions may also separate the dark matter from the plasma that can be explained without new physics, as is clearly evident in  extreme cases of the bullet cluster \citep{Mastropietro2008}, and  like the El Gordo  cluster \citep{Molnar2015}, where high speed collisions between pairs of clusters are ongoing. More direct evidence for collisional dark matter would be concluded from differences between the stellar and dark matter distributions as the stars behave like collisionless particles and we should expect the collisionless dark matter to follow the gravitational potential in the same way. Offsets between the position of the dark matter central peak  and the peak of the luminous matter are difficult to explain with a standard $\Lambda$CDM but are naturally produced for reasonable values of the dark matter cross-section \citep{Rocha2013}. A difference of this nature has been claimed recently based on detailed lensing data in the center of a cluster that contains 4 bright member galaxies \citep{Massey2015}. In the case of the Hubble frontier field clusters, it is interesting that our free form analysis of MACS0146 also indicates  a possible offset between the lensing based centroids of the brightest galaxies and their luminous stellar centroids. These differences are subtle and it will be important to look at a larger sample and the model dependencies, and systematic uncertainties, in detail to support any claim of new physics.

\begin{figure*} 
 {\includegraphics[width=16cm]{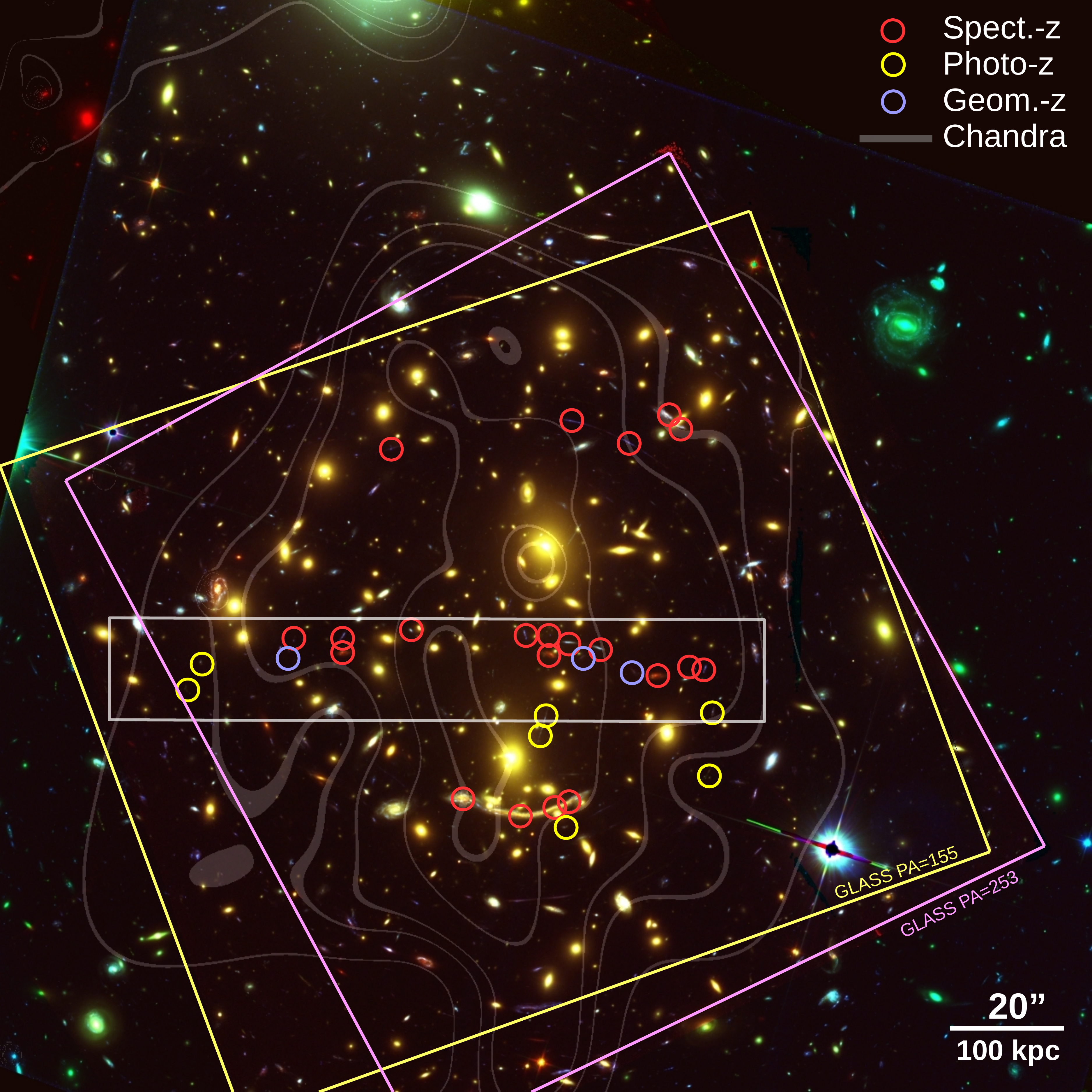}}
   \caption{A370 as seen by HST (red=IR bands, green = F814W band and blue = F415W+F606W bands) with {\it Chandra} contours overlaid. The field of view is 3.2\arcmin. The circles mark the positions of the multiply 
            lensed systems with spectroscopic, reliable photometric and/or reliable geometric redshift used to build the preliminary driver lens model.
            The gray central rectangular region marks the area with the highest density of reliable lensing constraints. The yellow and magenta square regions mark the field-of-view of the GLASS data at two position angles.}
   \label{fig_1}  
\end{figure*}

In this paper, we explore the cluster A370 \citep[$z=0.375$, ][]{Mahdavi2001} using our free-form code {\small WSLAP+} \citep{Diego2005,Diego2007,Diego2016,Sendra2014}. 
Our algorithm does not rely on assumptions on the distribution of the dark matter and 
seems to perform equally well in clusters that are more or less symmetric as clusters that present a complex morphology. 
A370, the final cluster to be completed in the HFF program (in September 2016), contains two very large elliptical galaxies with extended cD light profiles separated by roughly 190 kpc. The HFF data reveals that long radial arcs point towards each of these galaxies (although not quite precisely) suggesting that A370 is a double cluster with two overlapping cores that are clearly in the process of merging. The current X-ray data will be improved upon as part of the related HFF campaign, but currently possible substructure is not very significant in the relatively shallow X-ray images available at present which shows only that the generally elliptical distribution of member galaxies is similar in shape to the X-ray emission. 

This cluster has been studied intensively in the context of gravitational lensing since the discovery of cluster lensing based on the giant arc in this cluster \citep{Paczynski1987}, including the magnification of record breaking high redshift galaxies \citep{Hu2002}. Previous work on this cluster has identified a set of multiple lensed images with their redshifts and produced reliable lens models \citep{Johnson2014,Monna2014,Richard2014,Umetsu2015,Zitrin2015}. 
To date, $\sim 12$ multiply lensed galaxies have been reliably identified  \citep{Johnson2014,Richard2014} in the redshift range $0.8<z<6$. We can also make use of the recent redshift catalogs  compiled by Grism Lens-Amplified Survey from Space\footnote{https://archive.stsci.edu/prepds/glass/} \citep{Schmidt2014,Treu2015}  or GLASS hereafter, to examine  and in some cases update redshifts adopted in earlier work.  GLASS data is useful for both goals, confirming the cluster members 
and identifying new multiply lensed images. The data from the HFF and GLASS programs are complemented with X-ray data from Chandra to trace the hot baryonic component in the core of the cluster. 

The free-form nature of our method means we can objectively determine the relative distances of any set of multiple images with respect to any other, providing geometric estimates of the source distance for comparison with the independently determined photometric redshifts. We have applied this approach increasingly since it was first demonstrated in A1689 \citep{Broadhurst2005} because the accuracy of the free-form lensing is now sufficiently  reliable for this purpose given the relatively high surface density of multiply lensed images reached in the HFF and allows us to confidently measure our own geometric redshift estimates \citep{Lam2014} including photometrically ambiguous high redshift galaxies at $z\simeq 10$ \citep{Zitrin2014,Chan2016}.

This paper is organised as follows. In section~\ref{sect_S2} we introduce the HFF and GLASS data used in this study and briefly describe the X-ray data. 
In section~\ref{sect_S4} we present the initial lensing data used to constrain our preliminary (or {\it driver}) model. 
Section~\ref{sect_S5} describes the algorithm used to derive the lens models. 
Section~\ref{sect_S6} discusses the reference (or driver) model and alternative models that are presented in the paper.  
Results based on the driver model applied to new HFF and GLASS data and a discussion of the new systems that we are able to uncover 
using the driver model are presented in section~\ref{sect_S7}. 
In section \ref{Sect_Sims} we compare our results with simulations from the MUSIC project\footnote{http://music.ft.uam.es/}. 
Our results are discussed in section~\ref{Sect_Discus} and we conclude in section ~\ref{sect_S8}. 

Throughout the paper we assume a cosmological model with $\Omega_M=0.3$,
$\Lambda=0.7$, $h=70$ km/s/Mpc. For this model, $1\arcsec = 5.16$ kpc at the distance of the cluster ($z=0.375$). 
In all images (except when noted otherwise) we adopt the standard convention where north is up and east is left.

%%%%%%%%%%%%%%%%%%%%%%%%%%%%%%%%%%%%%%%%%%%%%%%%%%%%%%%%%%%%%
\section{HFF, GLASS, and X-ray data}\label{sect_S2}
%%%%%%%%%%%%%%%%%%%%%%%%%%%%%%%%%%%%%%%%%%%%%%%%%%%%%%%%%%%%%
 
\subsection{HFF.} We use the reduced public imaging data obtained from the ACS and WFC3 Hubble instruments, retrieved from the Mikulski Archive for Space Telescope\footnote{https://archive.stsci.edu/prepds/frontier/} (MAST). For the optical data (filters: F435W, F606W and F814W), we used the recently released data that includes the first 76 orbits of HFF data on this cluster  (ID 14038, PI. J. Lotz) plus 6 orbits from previous programs (ID 11507, P.I K. Noll and ID 11591, P.I J-P. Kneib). For the IR data, we used data collected in the HFF program ( 2 orbits in the filter F140W, ID 14038, PI. J. Lotz) as well as previous programs in the filters F105W (1 orbit, ID 13459, P.I T. Treu), F140W (3 orbits, ID 11108, P.I E. Hu and ID 13459, P.I T. Treu), and F160W (3 orbits, ID 11591, P.I J-P. Kneib and ID 14216 P.I R. Kirshner) totaling 92 orbits in all six bands.  From the original files, we produce two sets of colour images by combining the optical and IR bands. The first set is based on the original data while in the second set we apply a high-pass filter to reduce the diffuse emission from member galaxies and a low-pass filter to increase the signal-to-noise ratio of small compact faint objects. The second set is particularly useful to match colours in objects that lie behind a luminous member galaxy where the light from the foreground galaxy affects the colours  of the background galaxy.

\subsection{GLASS.} 
\begin{figure}  
 \centerline{ \includegraphics[width=9cm]{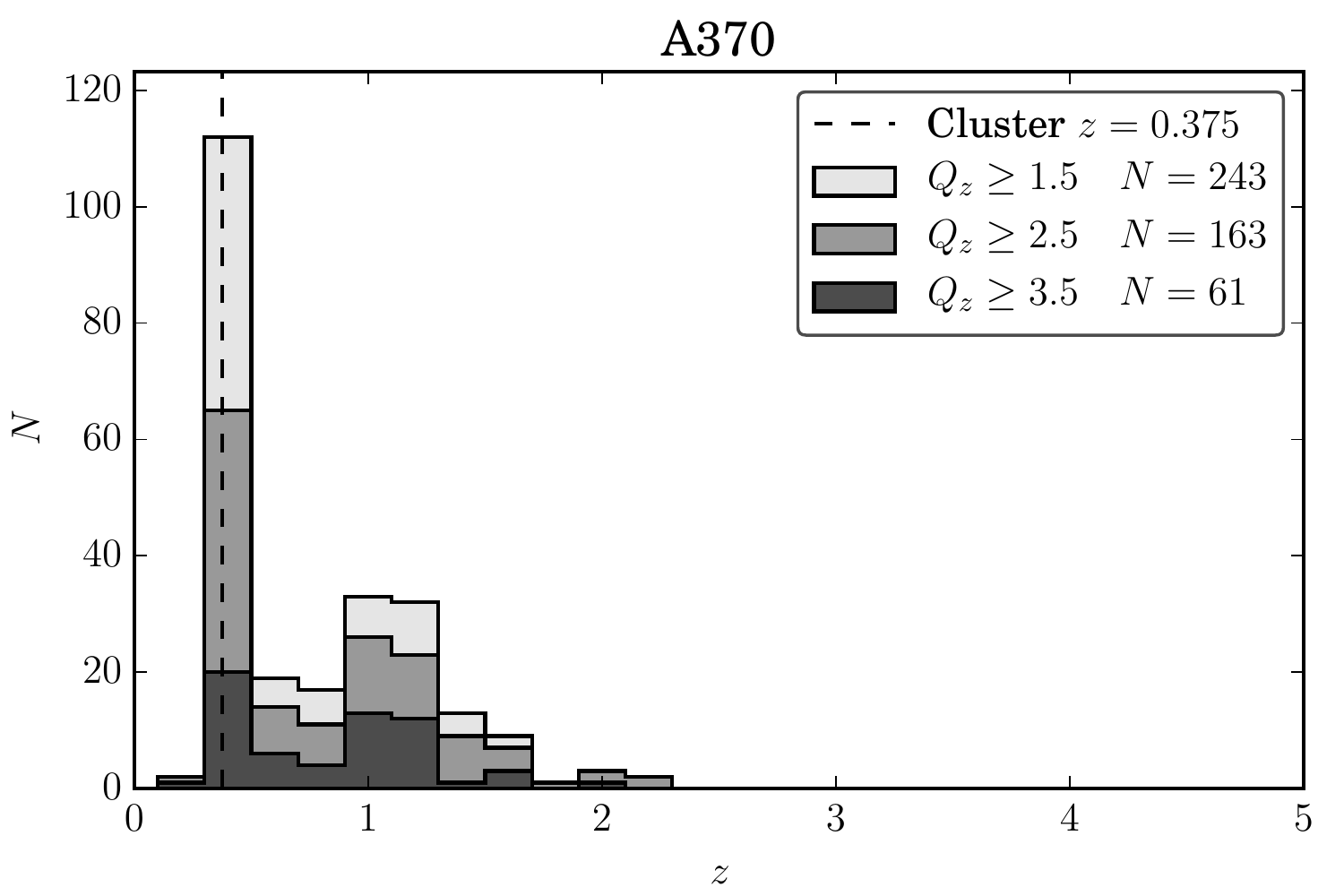}}  
   \caption{The redshift distribution of the publicly available GLASS v001 redshift catalog of A370. $Q_z$ refers to the quality of the individual redshifts ($Q_z=1$ worst; $Q_z=4$ best). Cluster members at $z\sim0.375$ (vertical dashed line) are clearly separated from the rest of the sample. A second over-density of objects is apparent at $z\sim1$. This overdensity is discussed in Section 8.  
           } 
   \label{fig_backgr_zx}  
\end{figure}

We make use of the spectroscopic redshifts of (multiply) lensed sources behind A370 as well as cluster members, including the publicly available GLASS (ID 13459, P.I T. Treu \citep{Schmidt2014,Treu2015} v001 redshift catalog available at \url{https://archive.stsci.edu/prepds/glass/}. %\url{https://archive.stsci.edu/prepds/glass/}. 
The GLASS data include HST grism spectroscopy in the WFC3 G102 and G141 grisms at 10 and 4 orbits depth, respectively, to obtain comparable 1$\sigma$ flux limits of $5\times10^{-18}$erg/s/cm$^2$ per position angle over the full wavelength range of the two grisms (0.8--1.7$\mu$m). Before each grism exposure, short pre-imaging in F105W and F140W was obtained for optimal alignment and extraction. These data are included in the imaging mosaics described above. The GLASS v001 redshift catalog of A370 was generated by careful vetting and visual inspection of the GLASS grism spectra of emission line sources and objects with a continuum H-band magnitude brighter than 23. As opposed to the majority of the GLASS v001 redshift catalogs, the A370 redshifts were determined without any photometric prior, as the HFF data were incomplete at the time the A370 redshift catalog was assembled. Each redshift was assigned a quality ($Q_z$)from 1 (worst) to 4 (best) corresponding to a redshift determined based on tentative low-S/N spectral features, and multiple high-S/N emission lines, respectively. For detailed general information on the GLASS v001 redshift catalogs see \cite{Treu2015}. The improved photometric information presented in the current study, has led to improved spectroscopic redshift estimates for a few objects as described in Section 4 and 5. The GLASS NIR footprint is shown in Fig.~\ref{fig_1} and the redshift distribution of GLASS sources is shown in Fig.~\ref{fig_backgr_zx}. The peak in the redshift distribution at z$\approx 1$ is discussed in section \ref{Sect_Discus}. 

\subsection{Chandra.} To explore the dynamical state of A370, we also produce an X-ray image using public Chandra data. In particular, we used data from the ACIS-I and ACIS-S instruments with the Obs ID 7715 and Obs ID 555 (PI. Garmire) totaling 85 ks. The X-ray data is smoothed using the code {\small ASMOOTH} \citep{Ebeling2006}. 
A false color image from the HFF imaging overlaid contours of the smoothed X-ray data is shown in figure \ref{fig_1}.
The distribution of X-rays seem to follow a smooth distribution with no obvious peak at the centre although some sub-structure may be present near the centre that shows some correspondence with the position of the BCGs and may become more clear with deeper planned X-ray data.

%%%%%%%%%%%%%%%%%%%%%%%%%%%%%%%%%%%%%%%%%%%%%%%%
\section{Lensing data}\label{sect_S4}
%%%%%%%%%%%%%%%%%%%%%%%%%%%%%%%%%%%%%%%%%%%%%%%%

For the lensing data we follow the multiple-image system identifications from \cite{Richard2010,Richard2014} and \cite{Johnson2014} that include between 9 and 12 multiply lensed systems (see compilation in Table~\ref{tab_arcs} in the appendix). From these papers we also adopt the numbering scheme (except for system 10 that is redefined in this work) as well as the spectroscopic redshifts. Some of the redshifts are updated with the recent estimates from GLASS (see section \ref{sect_S2}). 
We redefine image 8.3 in \cite{Richard2010,Richard2014,Johnson2014} which now is part of our new system 22. We find a new candidate for the third counterimage of system 8 (8.3) that is hidden behind one arm in a spiral galaxy that is itself lensed by a prominent member galaxy.
We also redefine system 7 with a new counter image, 7.5, that is incorrectly identified in \cite{Richard2010} and  \cite{Johnson2014}. Note that  \cite{Richard2014} did not include the original 7.5 from \cite{Richard2010} in their 2014 analysis. Table ~\ref{tab_arcs} compiles all systems (and their redshifts) used in this work. Support material including postage stamps of all system images are available at \url{http://www2.ifca.unican.es/users/jdiego/A370}. 

In addition to the centroid position of the multiply lensed systems, we can also use the position of individual knots present in the well resolved arcs, that are now much more readily identified between counter images thanks to the depth of the HFF data. In particular, systems 2, 3, 4, 5, 6, 7 and 16 contain distinguishing features that can be easily identified within their multiple images.  In the context of our free-form method, the addition of extra knots in well resolved systems greatly improves the accuracy and stability of the derived lensing solutions \citep{Diego2016} due to the large extension of the giant arcs. In addition, some of the elongated arcs that are not necessarily multiply lensed may be incorporated into the reconstruction  as additional constraints by requiring that these arcs should focus to a small region in the source plane. This additional information is especially useful in the regions beyond the critical curves where the density of multiply-lensed images drops. Systems 25 and 26 fall in this category.

%%%%%%%%%%%%%%%%%%%%%%%%%%%%%%%%%%%%%%%%%%%%%%%%%%%%%%%%%%%%%%%%%%%%%%%%%%%%%%
\section{Lensing reconstruction algorithm; WSLAP+}\label{sect_S5}
%%%%%%%%%%%%%%%%%%%%%%%%%%%%%%%%%%%%%%%%%%%%%%%%%%%%%%%%%%%%%%%%%%%%%%%%%%%%%%
We use our method WSLAP+ to perform the lensing mass reconstruction
with the lensed systems and internal features described above.
The reader can find the details of the method in our previous papers
\citep{Diego2005,Diego2007,Diego2016,Sendra2014}. 
Here we give a brief summary of the most essential elements. \\
Given the standard lens equation, 
\begin{equation} \beta = \theta -
\alpha(\theta,\Sigma), 
\label{eq_lens} 
\end{equation} 
where $\theta$ is the observed position of the source, $\alpha$ is the
deflection angle, $\Sigma(\theta)$ is the surface mass density of the
cluster at the position $\theta$, and $\beta$ is the position of
the background source. Both the strong lensing and weak lensing
observables can be expressed in terms of derivatives of the lensing
potential. 
\begin{equation}
\label{2-dim_potential} 
\psi(\theta) = \frac{4 G D_{l}D_{ls}}{c^2 D_{s}} \int d^2\theta'
\Sigma(\theta')ln(|\theta - \theta'|), \label{eq_psi} 
\end{equation}
where $D_l$, $D_s$, and $D_{ls}$ are the
angular diameter distances to the lens, to the source and from the lens to 
the source, respectively. The unknowns of the lensing
problem are in general the surface mass density and the positions of
the background sources in the source plane. 
The surface mass density is described by the combination of two components; 
i) a soft (or diffuse) component (parameterized as superposition of Gaussians) and 
ii) a compact component that accounts for the mass associated with the individual halos (galaxies) in the cluster. \\
For the diffuse component we find that Gaussian functions provide a good compromise between the desired compactness and smoothness  of the basis function. For the compact component we adopt directly the light distribution in one of the bands with the highest signal-to-noise ratio (F814W). 
To each galaxy, we assign a mass proportional to its surface brightness. This mass is later re-adjusted as part of the optimization process. 
The compact component is usually divided in independent layers, each one containing one or several cluster members. 
The separation into different layers allows us to constrain the mass associated to special halos (such as the giant elliptical galaxies) independently 
from more ordinary galaxies. This is useful in the case where the light-to-mass ratio may be different, like for instance in the BCG. \\

\begin{figure}  
 \centerline{ \includegraphics[width=8cm]{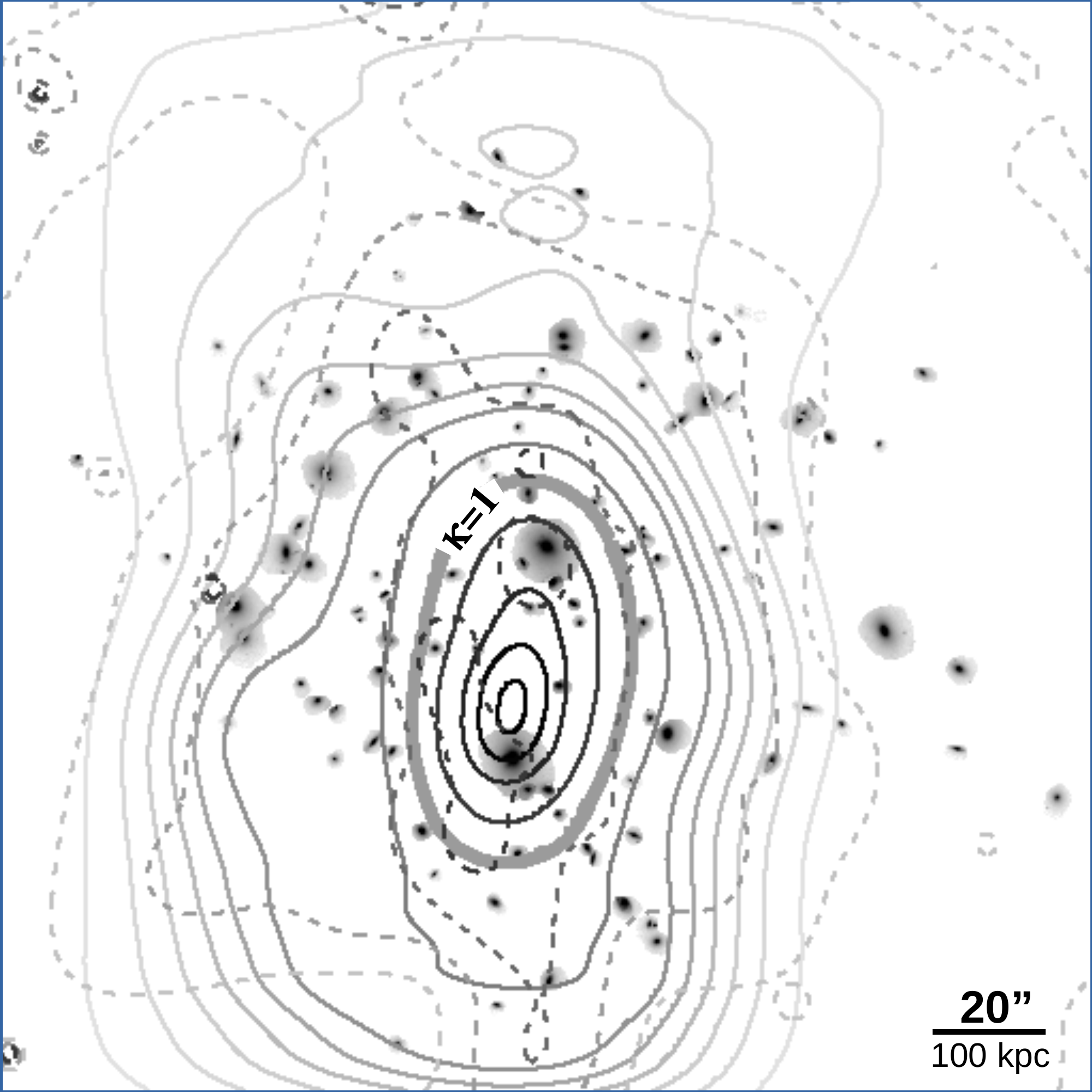}}  
   \caption{Contours of the grid component of the mass (solid lines) for the driver model compared with the X-ray contours from Chandra (dashed lines). 
            The thick solid line contour corresponds to $\kappa=1$ at $z=3$. 
            The galaxies used to model the small scale component are also shown. 
            Note how the peak of the dark matter from the grid component is very close to the region that is well constrained by the model.} 
   \label{fig_contour}  
\end{figure}

As shown by \cite{Diego2005,Diego2007}, the strong and weak lensing problem can be expressed as a system of linear
equations that can be represented in a compact form, 
\begin{equation}
\vectTheta = \matrGamma \vectX, 
\label{eq_lens_system} 
\end{equation} 
where the measured strong lensing observables (and weak lensing if available) are contained in the
array $\vectTheta$ of dimension $N_{\Theta }=2N_{\rm SL}$, the
unknown surface mass density and source positions are in the array $\vectX$
of dimension 
\begin{equation}
N_{\rm X}=N_{\rm c} + N_{\rm g} + 2N_{\rm s}
\label{eq_Nx}
\end{equation}
and the matrix $\matrGamma$ is known (for a given grid configuration and fiducial galaxy deflection field) and has dimension $N_{\Theta }\times N_{\rm X}$.  $N_{\rm SL}$ is the number of strong lensing observables (each one contributing with two constraints, $x$, and $y$), $N_{\rm c}$ is the number of grid points (or cells) that we use to divide the field of view. Each grid point contains a Gaussian function. The width of the Gaussians are chosen in such a way that two neighbouring grid points with the same amplitude produce a horizontal plateau in between the two  overlapping Gaussians. In this work we consider only regular grid configurations. Irregular grids are useful when there is a clear peak in the mass distribution, for instance when the cluster has a well defined centre or a single BCG. In the case of A370, there are two similarly bright "cD-like" galaxies, suggesting that the general mass distribution may be bi-modal with two similarly massive clusters that have recently merged to form A370. In these situations, the regular grid is a safer approach as it implicitly assumes a flat prior for the mass distribution. The default regular grid  used in this work has $N_{\rm c}=25\times25=625$ grid points. In addition to the grid, we model the small scale fluctuations around the cluster members. We take advantage of the GLASS data to select the most prominent galaxies in the cluster that lie within $0.36 \pm 0.06$ (the interval is defined to include prominent cluster members like the BCG in the north that has a lower $z=0.32$ as discussed below). Although some of these galaxies may not be dynamically linked with the cluster, given their redshift and mass their small scale effect should still be noticeable if they are close to the line of sight of one of the observed lensed images.  Among these,  we correct the GLASS redshift ($z=0.380$ with $Q_z=2$) of a bright foreground galaxy with a reliable photo-$z$ of $z_{\rm mean}=0.168$. The low redshift of this galaxy means it is intrinsically of relatively low luminosity and so unimportant for the lens model. The northern BCG in the cluster has a lower redshift in GLASS ($z=0.32$) but with a low quality flag ($Q_z=2$) due to a lack of emission lines. Independent photometric redshifts with two different codes BPZ \citep{Benitez2000} and EAZY \citep{Brammer2008} result in $z = 0.385$ and $z = 0.36$ respectively, in agreement with being a cluster member. Chandra data also supports this hypothesis since a small X-ray peak is found at the position of this galaxy suggesting a local gas density enhancement/gas cooling  (of the intracluster gas) around this galaxy. $N_{\rm g}$ (in Eq.~\ref{eq_Nx}) is the number of deflection fields (from cluster members) that we consider. 
\begin{figure*}  
 \centerline{ \includegraphics[width=16cm]{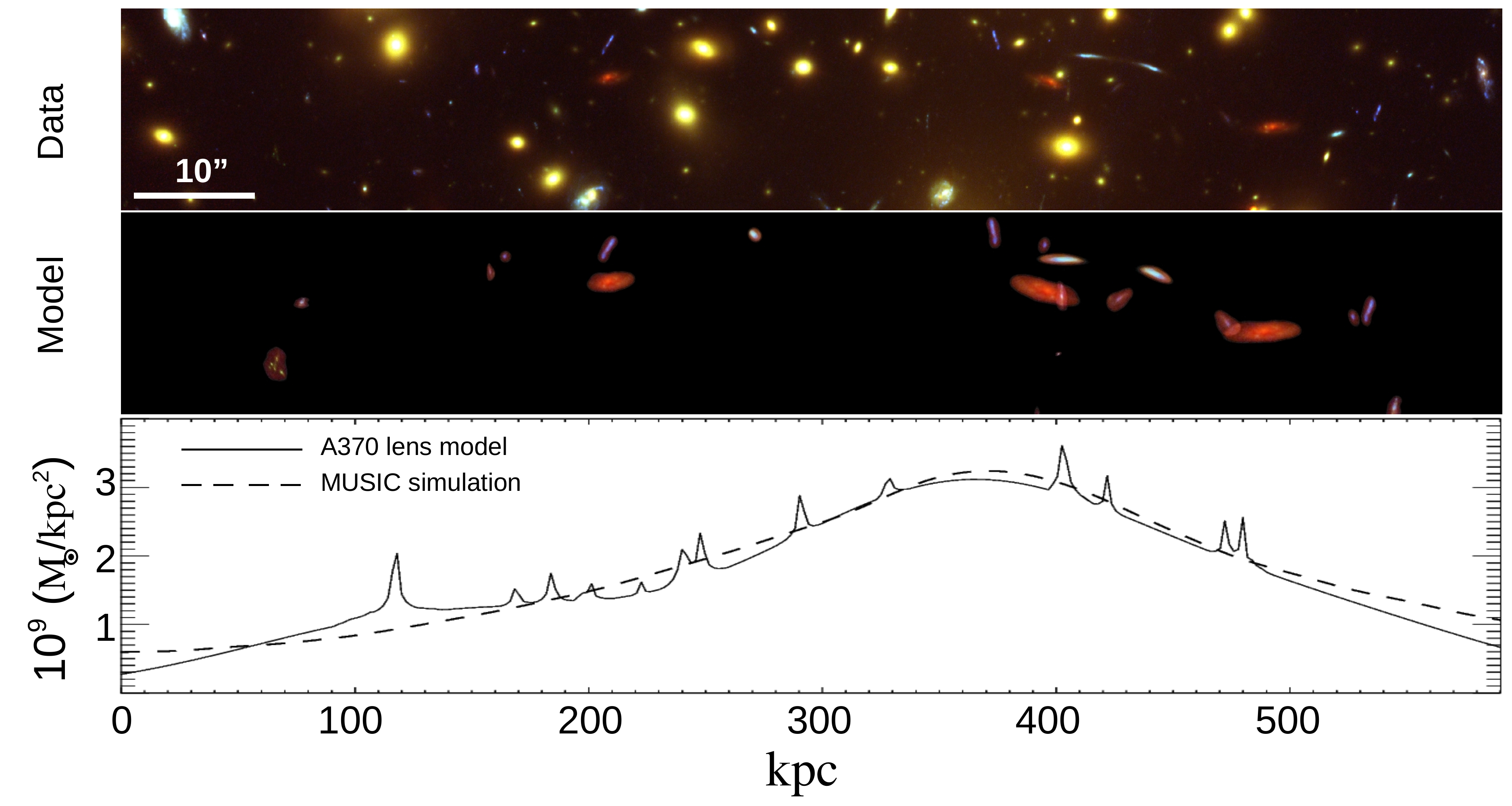}}  
   \caption{The top panel shows the rectangular sector highlighted in Fig.\ref{fig_1}. The middle panel shows the predicted lensed images based on the driver 
            model. The bottom panel shows the average total mass projected along the vertical direction within the rectangular region.} 
   \label{fig_2}  
\end{figure*}  
$N_{\rm g}$ can be seen as a number of mass layers, each one containing one or several galaxies 
at the distance of the cluster.  
In this work we set $N_{\rm g}$ equal to 1,2 or 3 to explore different configurations. 
In the case where  $N_{\rm g}=1$, all the individual galaxies in the lens model are assumed to follow the same light-to-mass ratio and are re-scaled by the same parameter (that is, they are all in the same layer). In a second scenario we assume $N_{\rm g}=2$ where all galaxies are in the same layer except the BCG that is in the southern part of the cluster (near the giant arc). The reason for this configuration is that the northern BCG seems to be poorly constrained by the lensing data so by adopting $N_{\rm g}=2$ we can explore the case where the mass-to-light ratio of the northern BCG is fixed together with the cluster members' mass-to-light ratio and we let the southern BCG be constrained by the lensing data. In the case where  $N_{\rm g}=3$, each BCG is allowed to have its own mass-to-light ratio and the remaining galaxies are placed in the third layer (and hence forced to have the same mas-to-light ratio). $N_{\rm g}=3$ results in models where the northern BCG contains a significantly larger mass (a factor $\approx 2$ larger) and predicts new arcs that are not observed in the data making this model less favored by the data. 
The particular configuration of the galaxies in our lens model is shown in figure \ref{fig_contour}. 

Finally, $N_{\rm s}$ in Eq.~\ref{eq_Nx} is the number of background sources (each contributes with two unknowns, $\beta_x$, and $\beta_y$) which in our particular case ranges from {\bf $N_{\rm s}=10$} when only the subset of reliable systems are used to {\bf $N_{\rm s}=30$} when all systems in Table~\ref{tab_arcs} are used in the reconstruction.The solution, $X$ of the system of equations \ref{eq_lens_system} is found after minimising a quadratic function of $X$ \citep[derived from the system of equations \ref{eq_lens_system} as described in ][]{Diego2005}. The minimisation of the quadratic function is done with the constraint that the solution, $\vectX$, must be positive. Since the vector $\vectX$ contains the grid masses, the re-normalisation factors for the galaxy deflection field and the background source positions, and all these quantities are always positive (the zero of the source positions is defined in the bottom left corner of the field of view), imposing  $\vectX>0$ helps constrain the space of meaningful solutions and to regularise the solution as it avoids large negative and positive contiguous fluctuations. The quadratic algorithm convergence is fast (a few minutes) allowing for multiple solutions to be explored in a relatively short time. Different solutions can be obtained after modifying the starting point in the optimization and/or the redshifts of the systems without spectroscopic redshift. A detailed discussion of the quadratic algorithm can be found in \cite{Diego2005}. For a discussion of its convergence and performance (based on simulated data) see \cite{Sendra2014}.

%%%%%%%%%%%%%%%%%%%%%%%%%%%%%%%%%
\section{Lens Models}\label{sect_S6}
\begin{figure*}  
 \centerline{ \includegraphics[width=18cm]{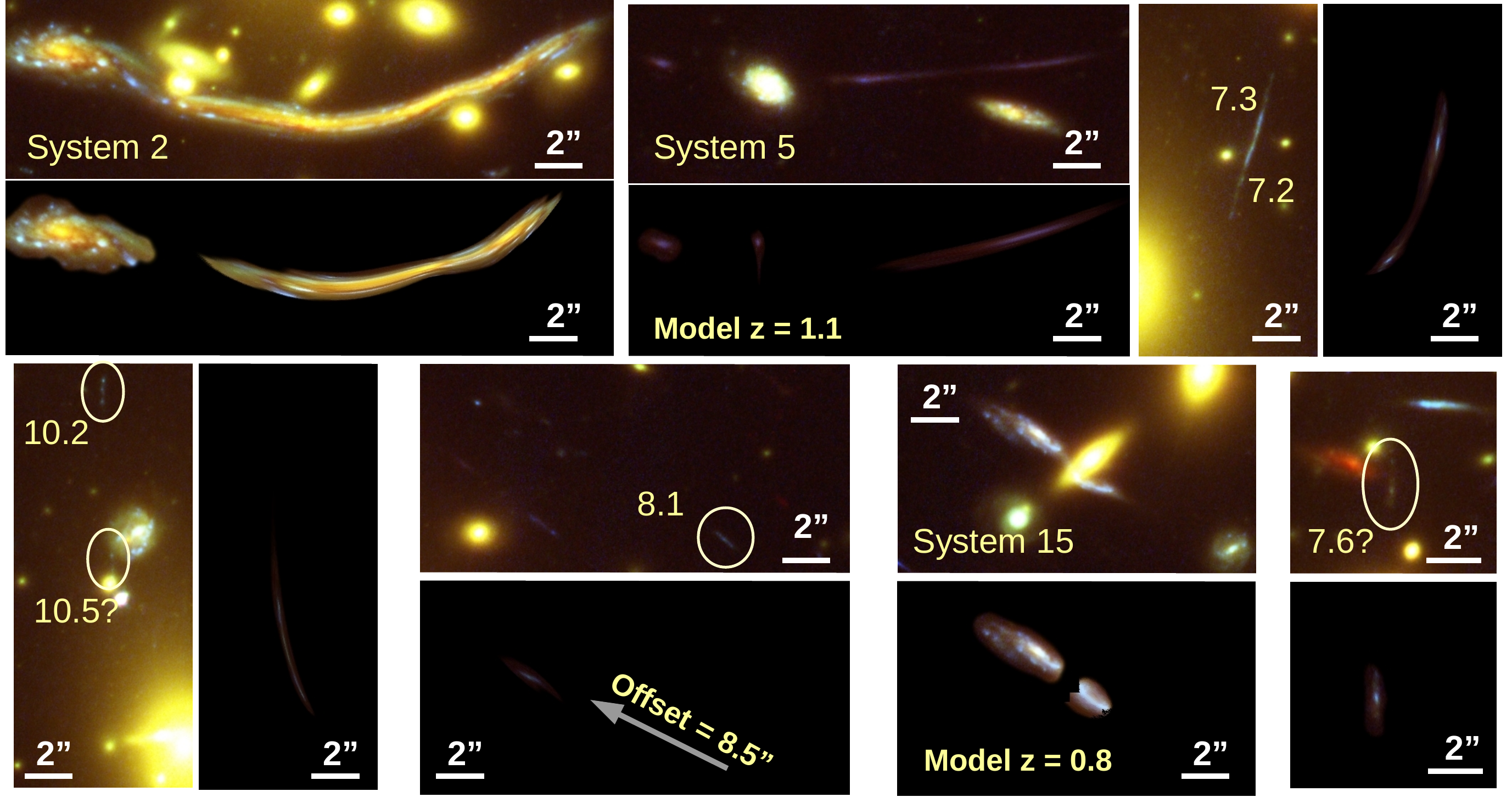}}  
   \caption{The driver model generally predicts the observed data with a high level of accuracy (Fig.~\ref{fig_2}). 
            This figure compiles some of the exceptions where the model does not perform as well (see text for a detailed discussion of each case). 
            The image candidate labeled as 7.6 is included in this figure is an example of the success of the model 
            given the accurate agreement of the appearance and location of this image with our prediction (note its somewhat redder colour is due to the incomplete current coverage of the IR data used in making this colour figure).} 
   \label{fig_SystemsIssues}  
\end{figure*}

%%%%%%%%%%%%%%%%%%%%%%%%%%%%%%%%%
The different lens models depend mainly on the assumptions made on i) the background sources and ii) the lens plane. In the following we discuss these two assumptions and how they impact the lens model. \\

\noindent
i) {\it Variability in the lens models linked to the definition of the background sources}. 
The assumptions made on the background sources refer to the number of multiply lensed systems and their redshifts used to constrain the lens model. A370 shows a number of multiply lensed systems. 
Among these, seven have reliable redshifts that make them easily identifiable as multiple images of the same source. Other systems 
however lack the redshift confirmation and are matched based on their morphology and colours (and photometric redshifts). 
Although most of the systems in the second group are probably multiple images of the same background source, the uncertainty in their redshift translates into an uncertainty 
in the lens model. This uncertainty can be reduced by restricting the analysis to the systems that are the most reliable. We identify ten such systems to which we assign rank A in table \ref{tab_arcs}. Most of these systems have robust spectroscopic redshifts. 
Among these, we update the redshift of system 3 with the new estimate 
from GLASS ($z_{\rm GLASS}$=1.95) from detection of the [OIII] doublet at $\sim 1.48$ $\mu{\rm m}$. Previously this system was assumed to be at redshift 1.42 \citep{Johnson2014}. Systems 7, 14 and 16, have no spectroscopic redshifts, however, their redshift estimate 
is very robust based on both photometric and geometric redshifts (from a lens model derived using only the systems with spectroscopic redshift). This robust set of 10 systems 
is used to derive a reliable preliminary lens model (the driver model) that is used to confirm/identify additional multiply lensed systems. For system 7, there is one candidate counterimage 
(7.6) that is not considered as part of the robust subset as it presents a different colour than the other counterimages. 
However, the colour variation may be a consequence of this counterimage being behind a very red object (system 6). 
On the other hand, the morphology and location of candidate image 7.6 agrees remarkably well 
with the prediction from the model (see Fig.~\ref{fig_SystemsIssues}).  
Both systems 14 and 16 are new systems that are confirmed with a high reliability by a version of the lens model 
that relies solely on the seven systems with spectroscopic redshift. A  very similar version of the lens model (that excluded also system 15 from the constraints) predicted system 15 with $z_{\rm geom}=1.55$ that was confirmed by GLASS data with $z_{\rm spect}=1.52$. We adopt the GLASS estimate for this system.  
Table \ref{tab_arcs} compiles the systems used to define the driver model and the systems that were either confirmed or discovered by the driver model. These systems can be included 
in the lens model to add more constraints in the reconstruction. Depending on what systems are included in the reconstruction the derived solution may vary. \\

\noindent
ii) {\it Variability in the lens models linked to the definition of the lens plane}. 
In our reconstruction, the lens plane is divided into two fundamentally different components. A soft component that is described by a grid of two-dimensional 
Gaussians and a compact component that is modeled following the light of the largest galaxies in the cluster. The grid component dominates over the compact component 
in terms of total mass and hence contributes most to the global deflection field. \\

Based on the differentiation made above, we consider three different scenarios or cases to derive the lens model

\noindent
$\bullet$ Case 1) The driver lens model. This model is based on the robust set of 10 systems described above having rank (A) 
in the column {\it Rank}  in table \ref{tab_arcs}. We assume a regular 
grid with $25\times25$ grid points and that all galaxies have the same light-to-mass ratio (i.e $N_{\rm g}=1$). \\

\noindent
$\bullet$ Case 2) Like above but using an extended sample with 9 additional systems listed in table  \ref{tab_arcs} as rank (B). 
These 9 additional systems are highly compatible with the lens model and have morphological features that increase their confidence. Some of these 
systems are useful to constrain the lens model in the regions where the high-confidence (rank A) constraints are scarce. \\

\noindent
$\bullet$ Case 3) Like above but using the full sample of 30 systems listed in table  \ref{tab_arcs} (ranks (A) (B) and (C)). Systems with rank (C) are in good agreement with the driver lens model but the lack of morphological features or more precise redshift estimates 
reduces their confidence with respect to systems having rank (A) and (B). \\

\noindent
$\bullet$ Cases 4,5,6) Like cases 1,2 and 3 respectively but allowing the southern BCG to have its own mass-to-light ratio (that is, we assume 
two layers for the compact component or N$_{\rm g}$=2). \\
 
\noindent
$\bullet$ Cases 7,8,9) Like cases 4,5 and 6 respectively but also allowing the northern BCG to have its own mass-to-light ratio, that is, we assume three layers for the compact component or N$_{\rm g}$=3. \\

Cases 1) through 9) are tabulated in table \ref{tab_cases}.
In addition to these models, in section \ref{sect_syst7} we introduce a new model, Case 10), to explain the radial arcs near the two BCGs. We refer to this additional model 
as the {\it shallow model}. 

\begin{table}
\centering
  \begin{minipage}{80mm}                                               
    \caption{Different cases assumed in the reconstruction of the lens models.  Each column corresponds to a different set of background sources. Rank A has the most reliable 
             systems, rank B is less reliable but still highly confident and rank C is the least reliable (but still consistent with the driver model). 
             N$_{\rm s}$ denotes the number of multiply lensed systems used in the 
             reconstruction.  N$_{\rm g}$ denotes the number of layers in the lens plane used to model the small component. All cases assume a uniform 
             grid with  N$_{\rm c}=25\times25=625$ grid points except for Case 10 (or {\it shallow model}) for which we use an adaptive grid. 
             This model (Case 10) is described in more detail in section \ref{sect_syst7}. }
 \label{tab_cases}
 \begin{tabular}{| c c c c |}   
 \hline
          &  N$_s$=10 (A)  &  N$_s$=19 (A,B) & N$_s$=30 (A,B,C) \\
 \hline
 N$_{\rm g}$=1  &   Case 1   &  Case 2    &  Case 3   \\
 N$_{\rm g}$=2  &   Case 4   &  Case 5    &  Case 6   \\
 N$_{\rm g}$=3  &   Case 7   &  Case 8    &  Case 9   \\
 N$_{\rm g}$=1  &   Case 10  &            &           \\
 \hline
 \end{tabular}
    \end{minipage}
\end{table}

%%%%%%%%%%%%%%%%%%%%%%%%%%%%%%%%%
\section{Results}\label{sect_S7}
%%%%%%%%%%%%%%%%%%%%%%%%%%%%%%%%%

Using the robust set of 10 systems discussed in section \ref{sect_S4} (and marked with rank A in table \ref{tab_arcs}) we derive the {\it driver} lens model. 
Seven of these robust systems are concentrated in a narrow rectangular area of less than 100 kpc in width between the two BCG (see Fig.~\ref{fig_1}).
The constraints in this part of the lens plane are therefore expected to be better than anywhere else. Fig.~\ref{fig_2} shows the accuracy of the driver 
lens model in this region by comparing the observed multiply lensed images (top panel) with the prediction based on the driver model (middle panel). The predicted 
images appear very close to the correct positions with errors $\sim 1''$. The morphology and orientation of all arcs is also well reproduced. 
The bottom panel shows the projection (along the vertical direction) of the surface mass density in the rectangular region. The most remarkable feature is the {\it plateau} in the smooth 
component between 100 kpc and 200 kpc in the bottom panel. On this side of the lens plane we find a group of prominent galaxies whose associated dark matter halo 
could be responsible for this plateau in the mass distribution. The peak in the soft distribution coincides with the line connecting the two BCG (see Fig.~\ref{fig_1})  
suggesting that each BCG is at the centre of either a nearly symmetrical halo or a halo that is oriented in the direction of the other BCG 
since the distribution of the dark matter in the rectangular region must be the result of the overlap of these halos. The orientation of one halo pointing towards 
the other is observed in N-body simulations owing to tidal forces and conservation of angular momentum \citep{Zhang2009},  originating from the tidal field due to the surrounding dark matter \citep{White1984}. 

The relative orientation of the two main halos can be inferred also from the 2-dimensional map of the surface mass density shown in Fig. \ref{fig_contour} in units of the critical surface mass density at $z=3$ (or $\kappa$). The soft component shows a clear alignment connecting the two BCGs. 
A "plateau" to the East is seen to extend towards a small group of galaxies in that region. Interestingly, the X-ray emission as observed by 
Chandra seems to indicate a small excess of X-ray emission (although with low significance) in this region. In previous work \citep{Lam2014} we found a correlation 
between the reconstructed surface mass density and the X-ray emission in regions relatively far from the cluster centre and where such a correlation was not anticipated. 
A small shift is observed between the location of the peak in the soft component of the dark matter distribution and the BCGs. This shift is worthy of a fuller dedicated investigation given its potential significance, but for the present we do not draw any firm conclusions.  

\begin{figure*}  
 \centerline{ \includegraphics[width=18cm]{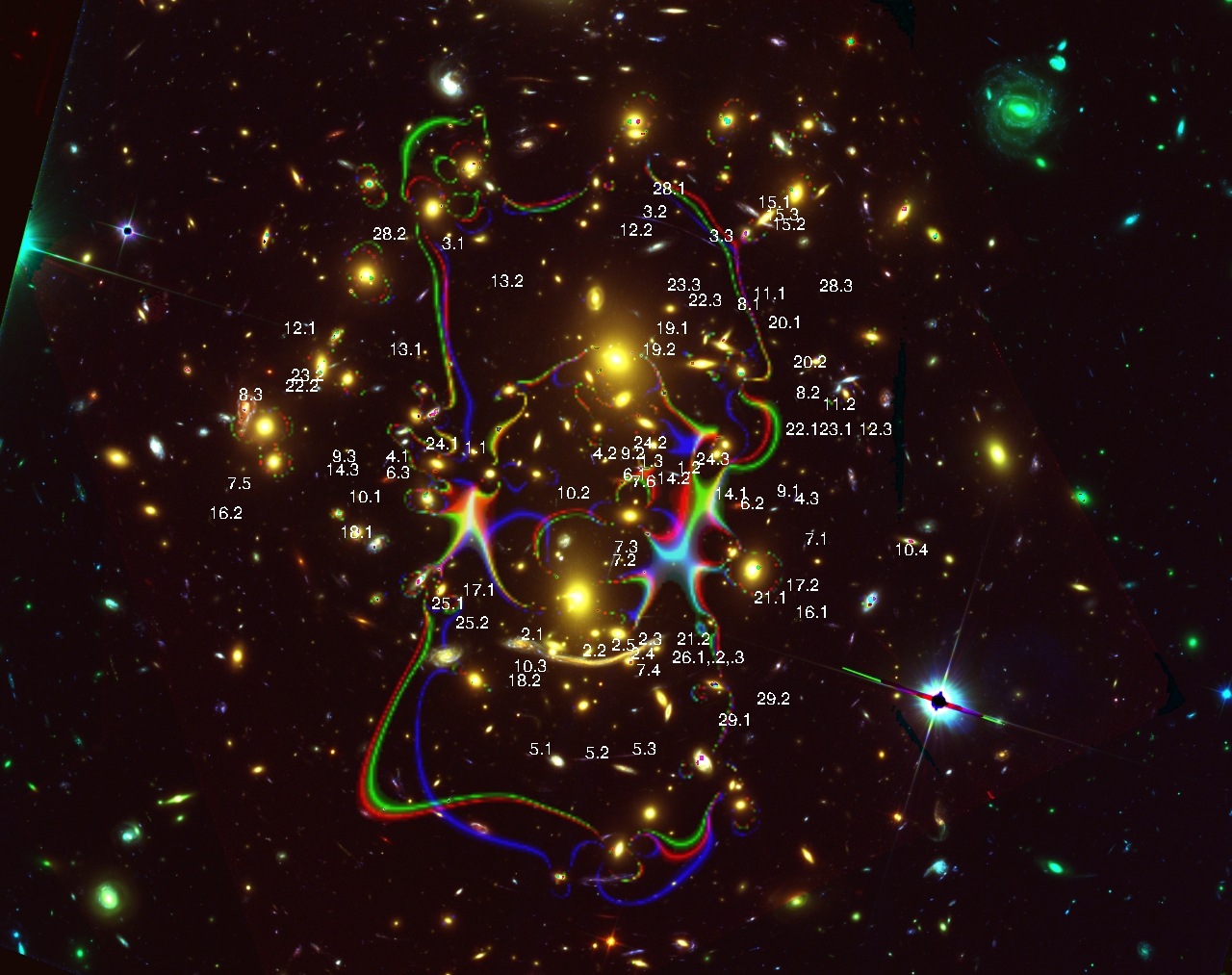}}  
   \caption{Critical curves for the Case 1 (blue), Case 2 (green), and Case 3 (red) lens model in Table~\ref{tab_cases} . The systems used in this work are marked in their location (with small shifts in some of them to avoid overlap). } 
   \label{fig_SystemsCritCurves}  
\end{figure*}

Based on the systems with rank (A) we vary the configuration in the lens plane by allowing the BCGs to have their own mass-to-light ratio (cases 4 and 7 in table \ref{tab_cases}). For case 4 we find that the southern BCG prefers a light-to-mass ratio that is similar to the rest of the galaxies. Hence, the solutions 
in cases 1 and 4 turn out to be very similar. For case 7 we find that 
the mass-to-light ratio of the southern BCG does not change significantly compared to the driver model. The northern BCG, on the other hand,  increases its mass considerably (by a factor $\sim 2$). This increase in mass results in additional predicted counterimages around the northern BCG that are not observed. 
Indirectly this is telling us that the dark matter in the vicinity of this BCG must have a softer (or shallower) core that is not well represented by the solution in case 7, which instead 
has a pronounced cusp at that position. The remaining cases in table \ref{tab_cases} imply the use of additional images. This is discussed in the next subsection.

%%%%%%%%%%%%%%%%%%%%%%%%%%%%%%%%%%%%%%%%%%%%%%%%%%
\subsection{New systems. Beyond the driver model}
%%%%%%%%%%%%%%%%%%%%%%%%%%%%%%%%%%%%%%%%%%%%%%%%%%

Using the driver model from case 1 above we confirm previously known systems as well as uncovering new candidate system with our model. The new systems are checked for consistency with photometric redshift estimates, and for consistency with 
the lens model in terms of morphology, parities and positions. The new systems that pass these checks are listed in table~\ref{tab_arcs} and shown in Fig.\ref{fig_SystemsCritCurves} and stamps of all systems are provided at \url{http://www2.ifca.unican.es/users/jdiego/A370}. 

For those  new additional systems that are not as secure as those above, we divide them into two categories. Rank (B) contains the systems that are highly reliable based on their morphology/parity but the redshift is more imprecise than in the systems with rank (A). Rank (C) contains system candidates that are highly compatible with the lens model but lack identifying morphological features (they are mostly nearly unresolved) and/or accurate redshifts prevents us from unambiguously identifying these systems. 
In most cases, the driver model predicts with a high level of accuracy the observed images. Some cases are not reproduced in such detail indicating deficiencies in the lens model. Such systems tend to group also in similar regions of the lens plane implying in these regions the driver lens model 
is less accurate. A few examples of the systems that are less compatible with the driver lens model are given in 
Fig.~\ref{fig_SystemsIssues}. Stamps of all the predicted images centered at the location of the observed images are also provided in the dedicated webpage. 
In this figure we see how systems that lie close to an overdensity of member galaxies, like system 2, makes modeling more challenging. Despite this, the driver model is able to reproduce the main features of system 2 and other systems. System 5 is predicted to be at $z\approx1.1$ by the driver lens model as opposed to $z>1.2$ inferred from photometric redshift. The presence of a spiral galaxy (a cluster member with $z=0.37$ and $Q_z=3$ from GLASS) between the counter images of the system, with a probable different light-to-mass ratio than the elliptical galaxies, makes this system hard to model without a specific model for the spiral galaxy.

 System 7 is of particular interest and will be studied in more detailed in section \ref{sect_syst7}. It has five robust counter images and probably a sixth one (shown in the bottom-right panel of Fig.~\ref{fig_SystemsIssues}) including an elongated radial arc (marked 7.2 and 7.3)  relatively close to the southern BCG composed of 2 images. We will also see that small changes in the lens model brings the new system 19 (similar in colour and morphology to system 7) into full consistency with being part of system 7, as described fully in section  \ref{sect_syst7}.  In detail, the radial arc (7.2,7.3) is not pointing directly to the BCG but to a position $\approx 15$ kpc west of the BCGs centre (the northern BCG has a similar elongated arc pointing also to a position  $\approx 18$ kpc west of the norther BCGs centre). The driver model predicts a curved shape for this radial arc, bending  towards the BCG (see Fig.~\ref{fig_SystemsIssues}) at odds with the straightness observed for this arc. The mismatch between the observed and predicted morphologies of the radial arc may be telling us that either there is a smaller amount of dark matter in the BCGs \citep[see for instance][to see how steep cusps suppress radial arcs]{Sand2002} or that the position of the peak of the dark matter is offset from the BCG which could have profound implications for the nature of dark matter. These offsets are predicted in simulations when the dark matter particles are allowed to interact with each other and exchange momentum \citep{Kahlhoefer2015}. Alternatively, if the total mass profile centered in the southern BCG is relatively shallow, then the minimum of the cluster potential need not coincide with the peak of the BCG mass distribution and hence offset with respect to the southern BCG. This is  be possible if the southern BCG has a significantly lower mass-to-light ratio than adopted in the driver lens model, as we shall explore in section  \ref{sect_syst7}. 

System 10 is highly consistent with the driver model except its central image (see bottom-left panel of Fig.~\ref{fig_SystemsIssues}). The driver model predicts a very elongated arc 
in the central region of the cluster that could be broken into several smaller arcs. 10.2 is in full agreement with the remaining counter images (in terms of colour and morphology). A possible 
new counter image 10.5 may be buried behind a member galaxy. System 8 (redefined in this work) is a robust system based on its morphology and colours. The driver lens model however predicts the 
location of one of the counter images at a position that is $\approx 8.5\arcsec$ from the position of the observed counter image. System 12 is also in the vicinity and has a similar offset indicating that the driver lens model lacks the accuracy in this region of the lens. The new system 15 shown in Fig.~\ref{fig_SystemsIssues} is well reproduced by the lens model 
when the redshift of the background source is assumed to be $z \approx 0.8$. The redshift measured by GLASS for this galaxy is $z=1.035$ ($Q_z=3$). In this case, the error is due to a larger mass assigned to the individual member galaxy. Since the deflection field scales linearly with the mass normalisation, assigning a mass 20\% smaller to this galaxy brings the predicted redshift to exact agreement with the measured value.

\begin{figure}  
% \centerline{ \includegraphics[width=10cm]{Mass_profile_A370.jpg}}  
 \centerline{ \includegraphics[width=10cm]{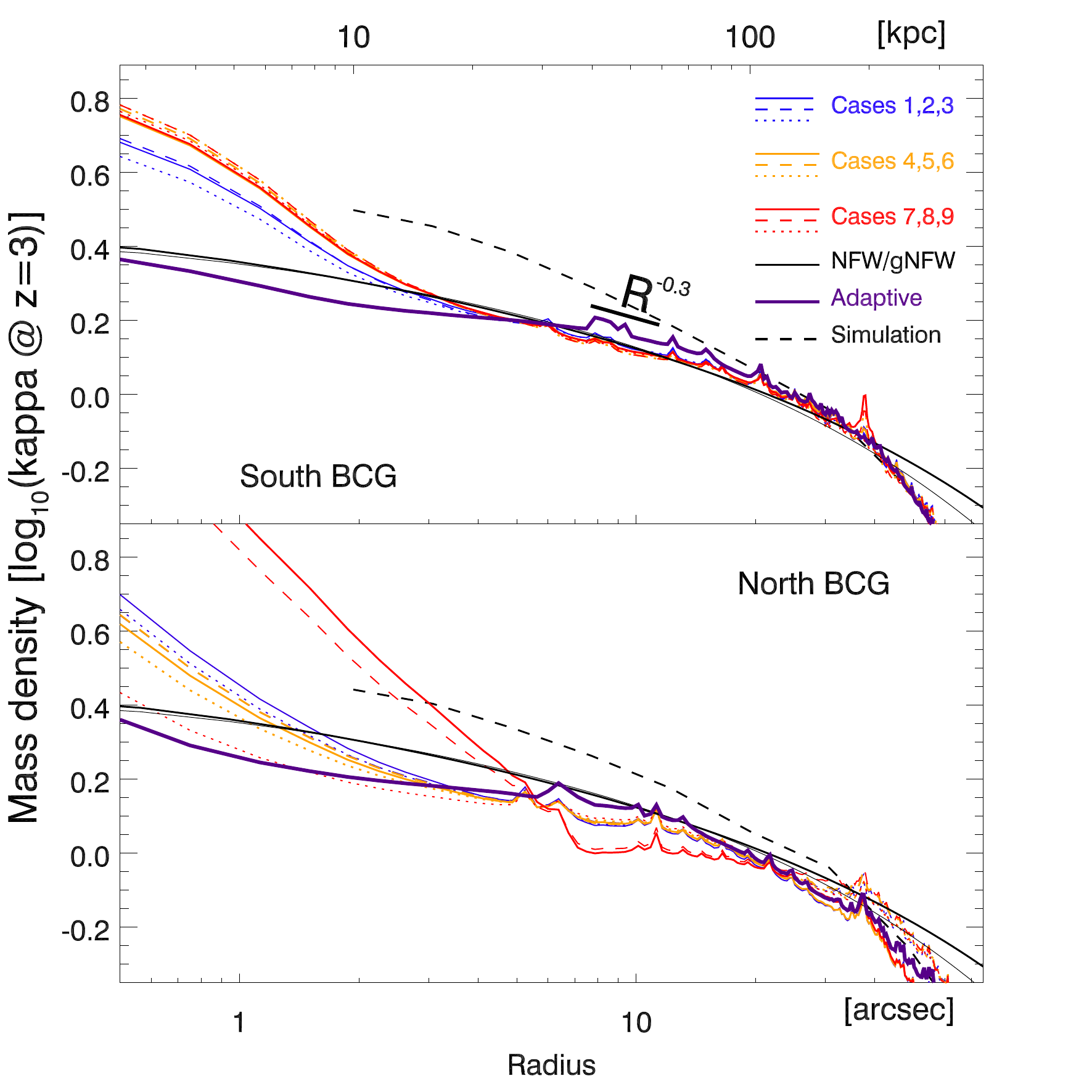}}  
% \centerline{ \includegraphics[width=10cm]{Mass_profile_A370.NEW.pdf}}  
% \centerline{ \includegraphics[width=10cm]{Mass_profile_A370.NEW.jpg}}  
   \caption{The total mass density (convergence) profile at $z=3$ for the models centered on the BCG in the south (top panel) and centered on the BCG in the north (bottom panel). 
            The coloured solid lines are for the cases 1,4 and 7 in table \ref{tab_cases}, the dashed lines are for cases 2, 5 and 8 and the 
            dotted lines are for cases 3, 6 and 9. The black solid lines correspond to an NFW model (thick solid line) and a gNFW model (thin solid line).  The NFW has a low concentration ($C=2$) and a virial radius = 3 Mpc. The gNFW model has the same virial radius but a larger concentration ($C=4$) compensated by a smaller inner slope ($\gamma=0.9$). The short solid line represents a power law ${\rm R}^{-0.3}$. The NFW and gNFW models are the same in both panels for comparison purposes. The thick dark blue solid line corresponds to the  
            alternative lens model discussed in section \ref{sect_syst7} (Case 10 in table 1). The black dashed line corresponds to the total mass (gas plus dark matter) profile of a simulation discussed in section~\ref{Sect_Sims}. The total mass  profile of the simulation below 10 kpc ($2\arcsec$) is not shown as it is below the smoothing length of the simulation.} 
   \label{fig_profiles}  
\end{figure}

Despite the apparent lower precision in the reproduction of the systems shown in Fig.~\ref{fig_SystemsIssues}, particularly the area around the northern BCG, 
the driver model performs remarkably well with most of the lensed systems, so here we make use of the driver model to seek new multiply lensed images  and to correct earlier work.  All the new systems that we uncover are shown in Fig.~\ref{fig_SystemsCritCurves} where we also show the critical curve of the driver model (in blue) 
and of two alternative models (cases 2 and 3). The critical curves show generally good agreement. The difference between cases 2 and 3 are smaller than the differences with the driver model (case 1). Cases 2 and 3 are part of the six additional lens  models (cases 2,3,5,6,8, and 9 in table \ref{tab_cases}) that we derive by varying the assumptions made on both the lens plane and the set of background sources.  

We compare quantitatively the nine models in table \ref{tab_cases} by computing radial profiles of total (i.e baryons plus dark matter) surface mass density for each model. Since the cluster contains two BCGs, we derive the profiles after centering on each BCG in turn.  
The results are summarized in Fig.~\ref{fig_profiles}  with the profiles around the south BCG in the top panel and around the north BCG in the bottom panel. 
For the BCG in the south the total mass density profile is relatively shallow between R=20 kpc and R=200 kpc and is well reproduced by an NFW profile with a very small concentration parameter of C=2. The shallowness and small concentration parameter are probably due to the overlap of the two cluster profiles at intermediate radii (R $\sim$ 100 kpc). Alternatively, the profile can be well reproduced by a gNFW profile with a larger concentration ($C=4$) but a smaller inner slope of $\gamma=0.9$ (the standard NFW profile has an inner slope of $\gamma=1$). In the range R$=[20-200]$ kpc all the solutions have very similar profiles. At small radii the profiles are more poorly constrained. 
For the BCG in the north, the reduced number of constraints translates into a larger variation between models, especially at R$<30$ kpc where the lensing constraints are 
the weakest. For cases 8 and 9, the northern BCG prefers a significantly larger mass ($\sim 2$ times larger) probably as a consequence of the addition  of new systems in the northern region 
that were not well reproduced by the driver model (like systems 8, 12 and 19).   
The profile in the bottom panel suggests also that the northern mass peak is slightly less massive than the southern peak, although both are very similar.   Note, in both cases (top and bottom panels), the second BCG appears in the radial profile as a bump at R$\simeq 200$ kpc, corresponding to the separation between the two BCGs.

\begin{figure}  
 \centerline{ \includegraphics[width=10cm]{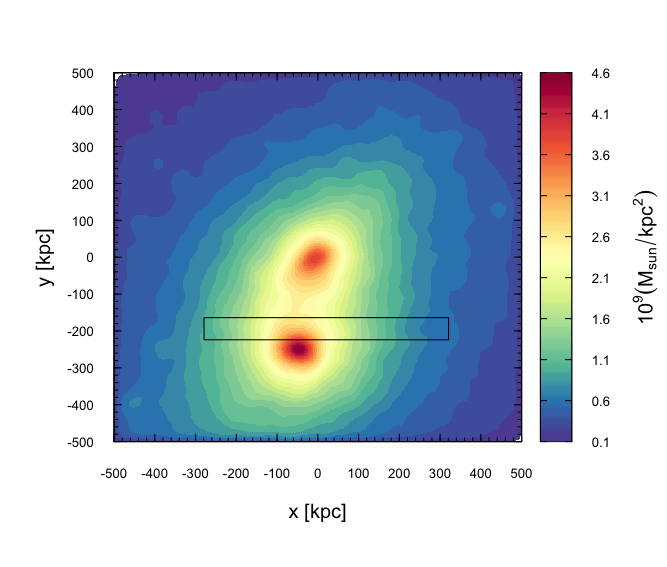}}  
% \centerline{ \includegraphics[width=9cm]{CL11H1p33_ADI_map_1LINE.pdf}}  
   \caption{Total mass density distribution in the MUSIC simulated halo at $z=0.333$. 
            The field of view is 1 Mpc$^2$ and the colour coding indicates the mass density in $10^9{\rm M}_{\odot}/{\rm kpc}^2$. 
            The rectangular sector marks the region in which we compute the sector projected mass shown in Fig.~\ref{fig_music2D}.
           } 
   \label{fig_music2D}  
\end{figure}  

%%%%%%%%%%%%%%%%%%%%%%%%%%%%%%%%%%%%%%%%%%%%%%%%%%%%%%%
\section{Comparison with simulations}\label{Sect_Sims}
%%%%%%%%%%%%%%%%%%%%%%%%%%%%%%%%%%%%%%%%%%%%%%%%%%%%%%%
We have extracted a cluster-size halo from the MUSIC data-set of re-simulated clusters
\citep{Sembolini2013}. In particular, we examined the non-radiative run (dark matter plus gas) of the MUSIC-2
data-set looking for a cluster-size halo resembling a similar projected mass distribution to that
of A370. The MUSIC-2 data-set constitutes a mass limited sample of cluster-like halos
selected from the low resolution version of the MultiDark 1 h$^{-1}$ Gpc simulation (\url{https://
www.cosmosim.org}). The MUSIC-2 halos have been re-simulated in spheres of 6
h$^{-1}$ Mpc radius centered in each cluster-size halo. Therefore, the mass resolution (or mass per particle) of the
MUSIC-2 halos is increased by a factor of 8 with respect to the parent simulation. After
projecting the halos along different lines of sight, we found an unrelaxed halo at $z = 0.333$
with two massive clumps separated $\sim 400$ kpc in projection. The spherical virial mass of the
halo is  M$_{\rm vir} = 1.25 \times 10^{15} h^{-1}{\rm M}_{\odot}$ and, for the projection selected, it produces an elongated
tangential critical line with an Einstein radius of  $\sim40\arcsec$ (for $z_s = 2.0$). 
It should be noted that (when projecting the halos) we only considered particles within a
parallelepiped of 6 h$^{-1}$ Mpc centered in the halo center. 
Figure \ref{fig_music2D} shows the projected mass
density as seen from an angle such that the apparent separation between the two clumps is
similar to that of A370. The rectangular region marks the sector over which we compute the projected 
profile (shown as a dashed line in Fig.~\ref{fig_2}). The projected profile in the rectangular region agrees remarkably well with the derived 
profile of the dark matter of A370 in a similar intermediate region (see Fig.~\ref{fig_2}). 
In terms of radial profiles, the simulated cluster resembles also the observed profiles at radii between 80 kpc and 300 kpc as shown in Fig.~\ref{fig_profiles}  (thick black dashed lines). 
At smaller radii, the profile from the simulation is steeper than the
observed one resulting in significantly denser central regions. 

%%%%%%%%%%%%%%%%%%%%%%%%%%%%%%%%%%%%%%%%%%%%%%%%%%%%%%%
\section{Discussion}\label{Sect_Discus}
\begin{figure}  
 \centerline{ \includegraphics[width=9cm]{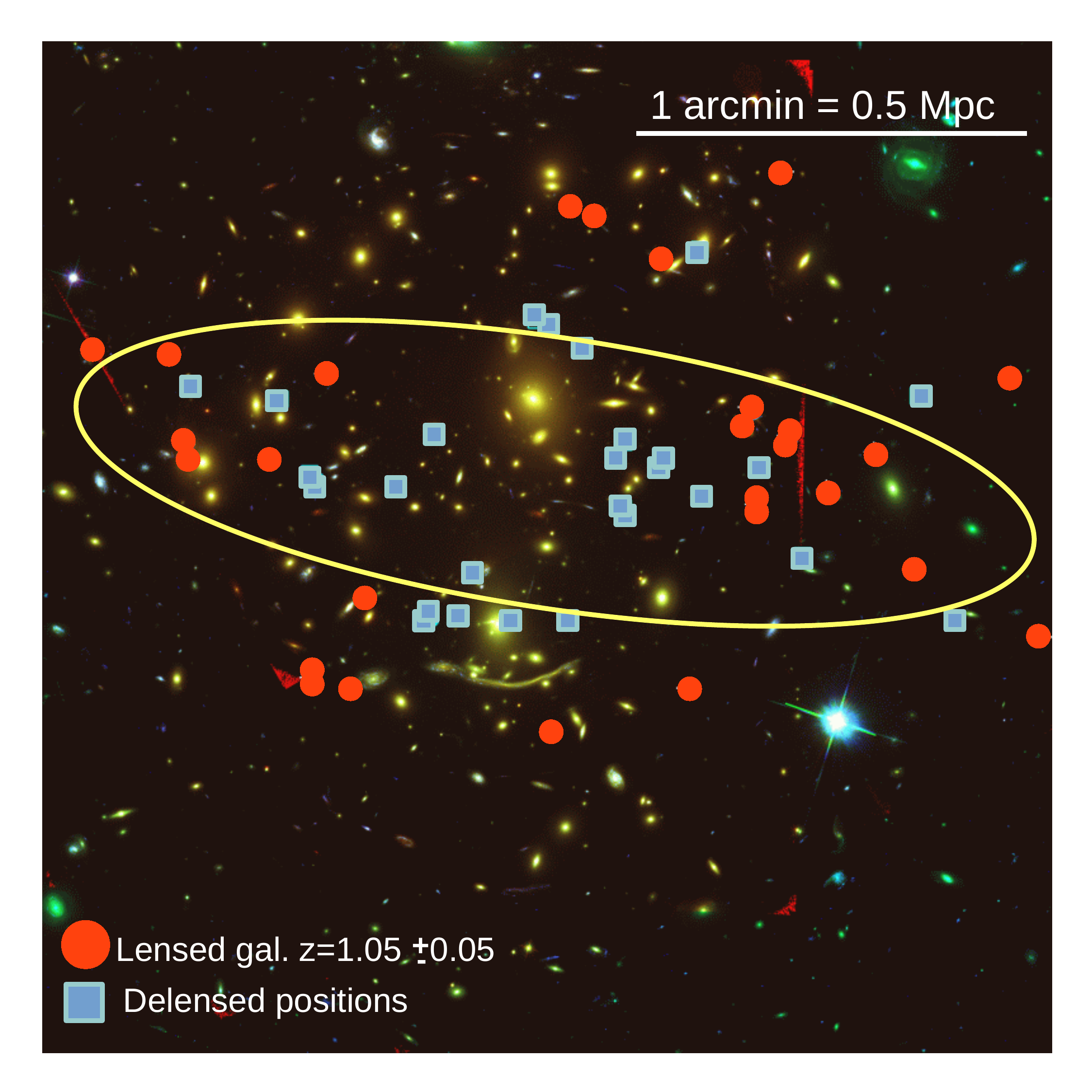}}  
   \caption{The red circles mark the observed positions of the galaxies in the narrow redshift interval z $\in [1.0,1.1]$ in the GLASS data. 80\% of the GLASS data in this interval have $Q_z>2$. 
            The squares mark the corresponding original positions in the source plane after deprojecting the observed positions with the driver lens model. 
            A filament-like structure (marked with a yellow ellipse) can be hinted closer to (and behind) the northern BCG.  
           } 
   \label{fig_backgr}  
\end{figure}

%%%%%%%%%%%%%%%%%%%%%%%%%%%%%%%%%%%%%%%%%%%%%%%%%%%%%%%
The models presented in the previous sections are generally able to reproduce the observed data. However, we observe some deviations that can give us useful information. As shown in Fig.\ref{fig_profiles}, the total mass distribution around the northern BCG is not constrained as well as around the southern BCG. This may be a consequence of the smaller number of constraints in the northern part. An other possible explanation may be projection effects that could be affecting the northern part of the cluster more than the southern part.The GLASS data, with its abundant redshift information, can be used to study the distribution of galaxies in the field of the cluster. Not surprisingly, we find that most of the GLASS galaxies are cluster members at $z \sim 0.37$. The second most prominent peak is at $z \sim 1.05$ (see Fig.~\ref{fig_backgr_zx}). When plotting the positions of the galaxies in this second peak, we find that they are not distributed in an homogeneous way but rather concentrated in a smaller region in the northern part of the cluster centre. The concentrated nature of these galaxies is made even more evident after we compute their original position in the 
source plane (see Fig. \ref{fig_backgr}). The galaxies at $z \sim 1.05$ seem to form a filamentary structure that lies behind the northern BCG of the cluster. It is 
reasonable to ask whether this structure at $z \sim 1.05$ may play a significant role in the lens model and if it does, then we may expect this impact to be larger in the 
northern part of the lens. More importantly, the impact should be larger on the higher redshift lensed systems at $z > 2$ found in the northern region while the central and southern region has more lensed systems with $z  \sim 1$ (or even $z < 1$) which would be unaffected by this structure. This structure may be the reason behind the relatively poorer performance of the lens model in the northern part of the lens.  
A second limitation of the lens models presented above is related to the reproduction of the systems near the BCG. Both BCGs have long radial arcs nearby that are not 
satisfactorily reproduced by the models in the previous sections. Here too, we can turn the problems that the lens models have in the central regions into an opportunity to gain 
some insight into the distribution of the dark matter around the BCGs. We do this by focusing on system 7 that is sensitive to the distribution of dark matter in the central 
region. 

\subsection{System 7}\label{sect_syst7}
%%%%%%%%%%%%%%%%%%%%%%%%%%%%%%%%%%%%%%%

\begin{figure*}  
 \centerline{ \includegraphics[width=18cm]{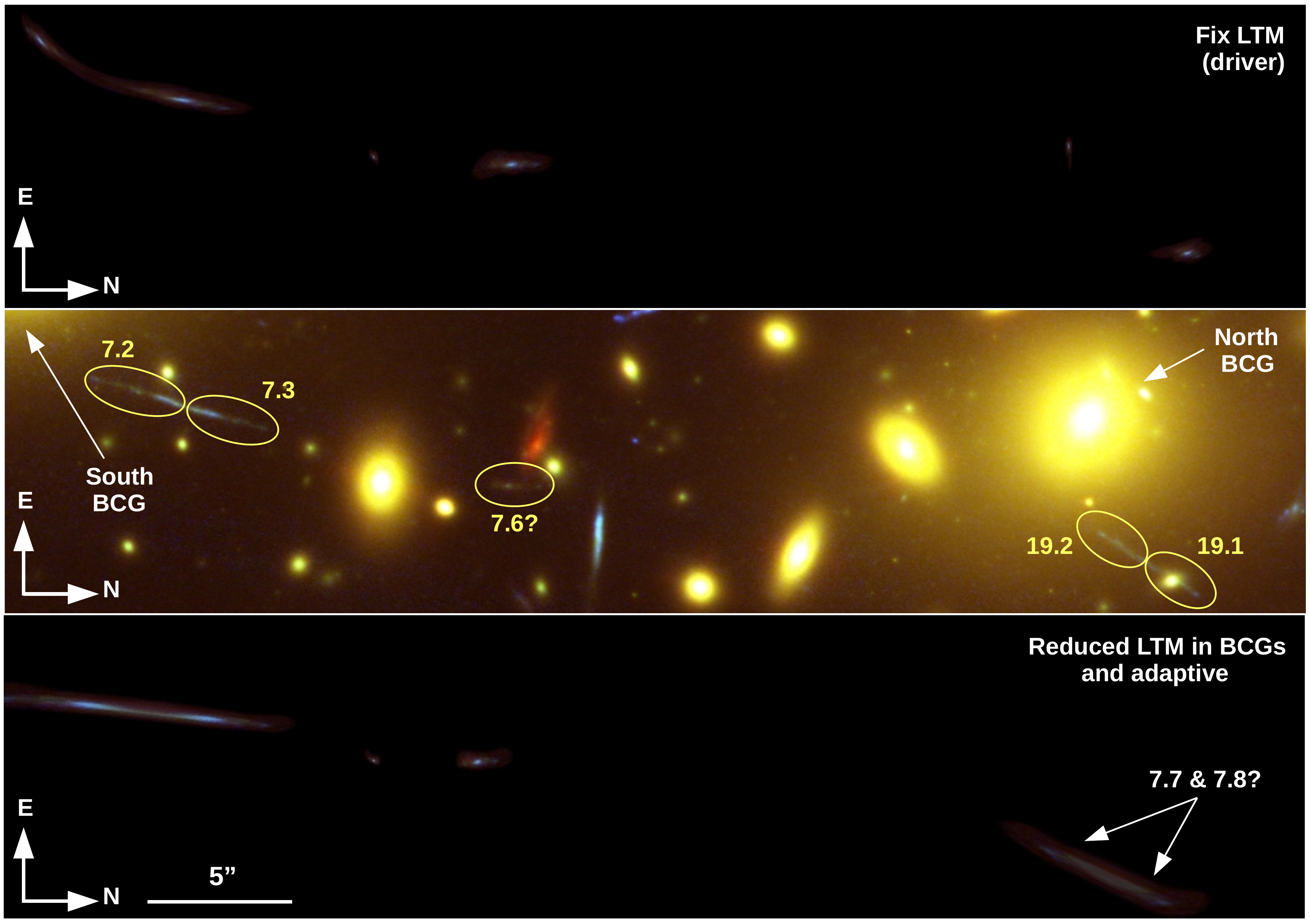}}  
   \caption{Central counterimages of system 7 as predicted by the driver model (top), as seen in the data (middle) and as predicted by the alternative multi-resolution model with reduced contribution from the BCGs (bottom). Note how the multi-resolution model straightens up the arc 7.2-7.3. The images are rotated $90^{\circ}$ (clockwise) with respect to the original image.} 
   \label{fig_Systems7}  
\end{figure*}

System 7 (and in particular the counterimages 7.2 and 7.3, see Fig.~\ref{fig_SystemsIssues}) is interesting for several reasons. The elongated radial arc in system 7 (7.2 and 7.3) 
confirms indirectly that the projected total mass profile is relatively shallow in that region of the lens \citep{Sand2002}. Its orientation points in a direction 
which does not coincide with the position of the BCG. As mentioned earlier, this could be explained if the mass around the southern BCG is relatively small and shallow. Also, the presence of other nearby compact clumps/galaxies may introduce small distortions in the potential around the BCG affecting the orientation of the radial arcs. The presence of nearby massive galaxies to the south-west of the BCG supports this possibility. Another interesting possibility is that the peak of the dark matter is not coincident with the BCG. As shown by N-body simulations, an offset between the peak of the dark matter and the position of the BCG is possible after a collision of two clusters if the dark matter particles have a certain probability of interaction with other dark matter particles. 

To test these two scenarios we produce a new model that minimizes the role of the two BCGs in the lens model and instead increases the resolution of the grid as it approaches each BCG. The new grid is adaptive  and doubles the resolution of the grid in the driver model (and the other models) near the position of the BCGs while gradually decreasing the resolution at larger distances (the resolution at the edge of the field of view is $\approx 75\%$ worse than in the driver model). The number of grid points in the adaptive grid is reduced by almost 50\% with respect to the driver model (lowering the number of degrees of freedom of the lens model). The mass-to-light ratio of the two BCGs is set to 20\% of the value used in the driver model. As in the driver model, we also consider only one layer, thus forcing the two BCGs to adopt a secondary role in the minimisation. Finally, as in the case of the driver model, for Case 10 we use only the systems with rank (A). 
In summary, the settings of the Case 10 model are similar to the ones for the driver model except for the grid (adaptive for Case 10 and regular for the driver model), and the light-to-mass ratio of the BCGs.
The increased resolution of the grid around the BCGs should produce an alternative model to those presented
above, but where the mass distribution in the central region has more freedom to change to accommodate the observations. The solution obtained with the multi-resolution 
model resembles the previous models with small differences. One of these differences is, as expected, around the BCGs where the new model is shallower 
than the Case 1--9 models in table \ref{tab_cases}. The profiles are shown in Fig.~\ref{fig_profiles} as a solid thick dark-blue curve. The position of the peaks in the dark matter are still consistent with being coincident with the BCG with no obvious offset, but the amplitude of the total mass at the position of the BCGs is smaller. 
We find that the multi-resolution model (Case 10) predicts a straight radial arc, whereas the driver model 
produced a curved radial arc (see Fig.~\ref{fig_Systems7}). The arc predicted by the  multi-resolution grid points towards another prominent galaxy to the south-west of the BCG  for which the mass-to-light ratio is set to the same value as in the driver model. 
The observed arc lies somewhere in between the predictions made by the driver model and the multi-resolution model so 
it is reasonable to assume that the profile of the true underlying mass is also somewhere in between the profiles of the driver model and the multi-resolution model. Both these models predict an additional counter image for system 7 that matches the position and morphology of the image labeled as 7.6 in Fig.~\ref{fig_SystemsIssues} and Fig.~\ref{fig_Systems7}. 
In addition, the multi-resolution model demonstrates that system 19 actually belongs to system 7. The morphology and location of the pair of radial images of system 19 are accurately predicted. It is important to realise that only part of the image of system 7 is being multiply lensed here, that does not include the bright blue central part of the source as the caustic
which bisects the radial image 19 includes only one end of the source. The detail with which we can reproduce this
system 19 leaves us in little doubt about its membership  of system 7. The driver model hinted already at this possibility of the relation between system 7 and 19 but  although a counter radial counter image is predicted at the location of system 19, the match in shape is  not accurate, as shown in  ~\ref{fig_Systems7}. The multi-resolution  grid, with its shallower less massive  BCGs predicts elongated arcs in detailed agreement with the data. Spectroscopic confirmation that systems 7 and 19 have the same redshift would clearly then support the lower mass profiles for the BCGs
favoured by the multi-resolution model. 
 
%%%%%  XXXX Kasper -->
We inspected the GLASS data at the positions of systems 7 and 19.  
For system 7, GLASS data reveal a tentative line in three of the counterimages of this system at around 14000A. This would correspond to a redshift $z\sim2.75$ for system 7 for the (unresolved) [OII] doublet. 
System 19 does not show any lines in the spectra. Photometric redshitfs derived for system 19 using the codes 
BPZ and EAZY result in $z_{\rm BPZ} \sim 0.3 $ and $z_{\rm EAzY} \sim 0.6$ respectively. For these redshifts, the lens models predict no radial arcs 
at the position of system 19 suggesting that the photometric redshifts may be affected by the nearby BCG. 
%%%%%  XXXX <--Kasper 

\subsection{BCG stellar mass and dark matter profiles }\label{sect_DMdepletion}
%%%%%%%%%%%%%%%%%%%%%%%%%%%%%%%%%%%%%%%%%%%%%%%%%%%%%%%%%%%%%%%%%%%
Inspired by the multi-resolution model discussed in the previous subsection, we take this as the best representation of the projected mass around the two  BCGs and we now estimate the contribution of the stellar mass to the lensing mass profiles shown in Fig.~\ref{fig_profiles}. The stellar mass distribution of the two BCGs is directly taken from the light profile using GALFIT, where all bright objects around the two BCGs are masked in making a 2D fit. For each BCG we use  a double Sersic profile to model their cores and extended light profiles. With this BCG light profile, we can convert it into a stellar mass distribution by using the observed spectral energy distribution (or SED) of each BCG.  %using the stellar population synthesis package, FAST (Kriek et al. 2009). 
Our SED fitting procedure includes the following: (i) the \cite{Bruzual2003} stellar population synthesis model, (ii) the \cite{Chabrier2003} initial mass function, (iii) an exponentially declining star formation history, (iii) the \cite{Kriek2013} dust law, (iv) a fixed redshift of $z=0.375$, and (v) solar metallicity $Z=0.02$. 
For the remaining parameters (e.g. star formation timescale, age, extinction, etc.) we sample a range of values with Monte Carlo to cover a fuller range of possible parameter choices in this context. 
The confidence levels are estimated by performing 100 Monte Carlo simulations.
We obtain a best fit value for the  total stellar mass of both BCGs combined of $M_{\rm BCG}= 1.15 \pm^{1.73}_{0.17} \times 10^{12}$ M$_{\odot}$ (3-$\sigma$ interval). 
We use the above set of solutions for the BCG light profiles together with a simple parameterization for the dark matter in a joint fit to the mass profile, to provide an underlying distribution of projected dark matter to associate with the two BCGs. For this we adopt a model with a flat core and a scale radius of this form: 

\begin{equation}
\kappa = \frac{1.75}{1 + \frac{R}{160 {\rm kpc}}}  
\end{equation}
for the southern BCG and a slight lower normalisation
for the northern BCG:
\begin{equation}
\kappa = \frac{1.60}{1 + \frac{R}{160 {\rm kpc}}}  
\end{equation}
This is because we aim to obtain a total mass that approximately matches our lens model profiles
above from the multi-resolution modeling, as the projected mass distribution from lensing of course
is a total mass including stars and dark matter. 

These models are shown as dotted lines in Fig.~\ref{fig_stellarmass}.  The combined profile of the total mass, from our dark matter profile and stellar mass profile (from GALFIT) matches very well the observed lensing profile from the multi-resolution model described in section 8.1 for both BCGs, as shown in Fig.~\ref{fig_stellarmass}. This figure demonstrates that within the $\sim 50$ kpc radius, where the stellar mass profile is derived, stars account for the lensing mass with little contribution from dark matter in either of the  BCGs. 

We should note that the discussion above is based on a Chabrier IMF. If a Salpeter IMF is assumed instead, the total stellar mass increases to  $M_{\rm BCG}= 2.00 \pm^{3.37}_{0.48} \times 10^{12}$ M$_{\odot}$ (3-$\sigma$ interval) making the argument above even more stringent. 
There is evidence that the IMF in massive galaxies is perhaps closer to Salpeter than to Chabrier \citep{Treu2010,Cappellari2012,Conroy2012,Newman2013b,Newman2015}. A Salpeter initial mass function could be still accommodated (while maintaining a constant density of dark matter in the central region) but would require that the total mass profile is indeed in between the driver model and the muti-resolution model as suggested by Fig.~\ref{fig_Systems7}. 

\subsection{Possible interpretations }\label{sect_Interpretation}
%%%%%%%%%%%%%%%%%%%%%%%%%%%%%%%%%%%%%%%%%%%%%%%%%%%%%%%%%%%%%%%%%%%
The  accurate model that we find for the radial arcs near the two BCGs perhaps provides one of the best constraints to date on the mass profiles of BCGs. Such detailed information is scarce and usually restricted to rare counter images that are small by comparison, so that the radial profile is not uniquely constrained \citep{Gavazzi2007}. Radial arcs have been used in the past to infer the inner slopes of the mass profiles in galaxy clusters and they are normally linked to shallow profiles in the centre of clusters \citep{Miralda-Escude1995,Bartelmann1996,Sand2002}. 

\begin{figure}  
 \centerline{ \includegraphics[width=9cm]{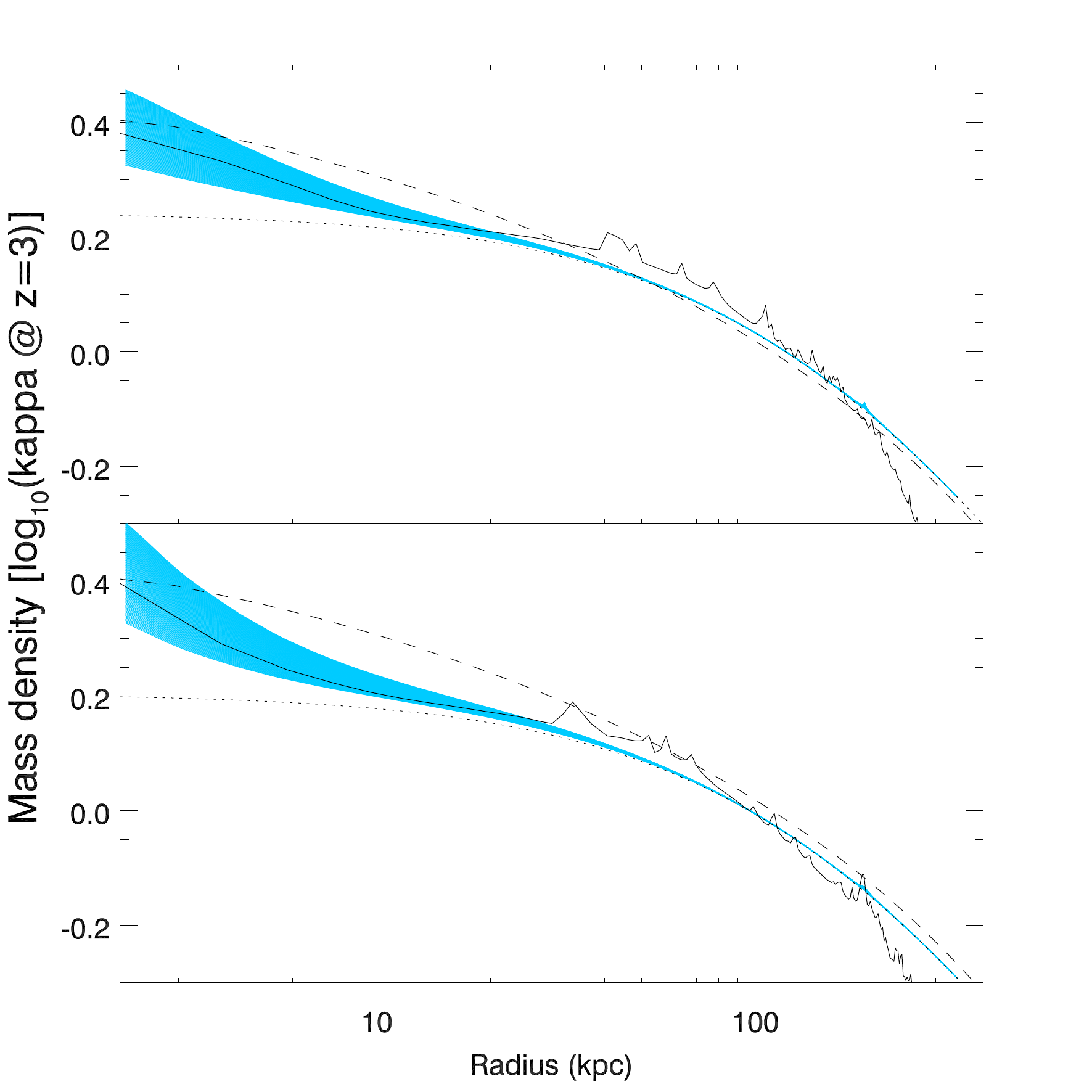}}  
   \caption{Stellar mass contribution to the BCGs. The solid lines correspond to the alternative model discussed in section \ref{sect_syst7} 
            (shown also as thick dark blue solid lines in Fig.~\ref{fig_profiles}). The top panel shows the main BCG in the south and the bottom panel shows the 
            secondary BCG in the north.
            The blue shaded region is the profile ($3\sigma$ interval) from adding the 
            estimated stellar mass to a dark matter toy model with a flat core (dotted lines). For comparison purposes we also show as dashed lines the NFW profile 
            (with concentration parameter $c=2$) of Fig.~\ref{fig_profiles}. 
           } 
   \label{fig_stellarmass}  
\end{figure}

If the multi-resolution solution is confirmed in the future (by independent lensing models or by confirming that systems 7 and 19 are  the same system), this would imply that the density of dark matter flattens inside the BCGs in contrast with predictions from N-body simulations although the total mass (including the stellar component) may be in better agreement \citep{Newman2013a,Newman2013b}. Simulations show that flattening of the central dark matter slope (below the canonical slope $\gamma=1$ from the NFW profile) is possible but only at distances below half-light radii of the BCGs \citep{Laporte2015}. 
BCGs typically present a concentration of dark matter that grows towards their centres. The projected constant density of dark matter inferred 
from Fig.~\ref{fig_stellarmass} suggests that these galaxies have a very small amount of dark matter inside them with a dark matter density similar to the density observed outside these galaxies. 
Scouring by supermassive black holes or ejection of the central supermassive black hole (or SMBH) has been proposed as one 
mechanism to flatten the inner light profile of massive galaxies \citep{Postman2012b} (and dark matter), but this will only temporarily affect the inner few kpc at most. Binary black holes can eject stars through three-body interactions, resulting  in flat cores in the luminosity profiles as  stars are also ejected (however we do not see evidence here of any particularly flat core in the stellar light profile). 
 Perhaps the most extreme case is the BCG in Abell 85 \citep{LopezCruz2014,Bonfini2015} with the largest known stellar core 
($\sim 5$ kpc) and that may be hosting a SMBH with M $\sim 10^{10}$ M$_{\odot}$ though  \cite{Madrid2016} discovered a small nucleus inside this BCG that may challenge this interpretation. Self-interacting dark matter is known to reduce the slope in the central region of clusters beyond 15 kpc \citep{Rocha2013,Harvey2015,Kahlhoefer2015}. For cross-sections $\sigma/{\rm m} \sim 1  {\rm cm}^2/{\rm g}$,  \cite{Rocha2013} finds that in haloes of masses ${\rm M}_{\rm vir} = 2 \times 10^{14}$  M$_{\odot}$  dark matter profiles flatten with core scales of $\sim 150$ kpc. 

The success of collisionless dark matter more generally means we may seek a less exotic interpretation involving gas cooling and the formation of stars within galaxy clusters as has been debated for many years but never conclusively shown.
Other mechanisms such as AGN feedback \citep{Martizzi2013}, supernova explosions \citep{Pontzen2012} or core heating by infalling baryons followed by a transfer of orbital energy to the dark matter particles \citep{ElZant2004} have been discussed as possible mechanisms. Perhaps the least explored possibility is sporadic AGN feedback that drives violent bulk gas motions \citep{Peirani2008} which heat the central dark matter, in analogy to the SN-driven bulk motions responsible for shallow cores in dwarfs. AGN feedback triggered by a SMBH is an interesting possibility as both BCGs are likely to host such SMBH. Chandra data reveals a small X-ray source in the northern BCG that could be a sign of AGN activity. The BCG in the south shows no evidence of associated X-ray emission. As discussed in for example \citep{Martizzi2013},  gas gets ejected by the AGN and returns after cooling.  This cycle generates gravitational potential fluctuations that modify the dark matter mass profile, resulting in a depletion in the galaxy of vast amounts of baryons and dark matter from the BCG. This mechanism works up to distances of $\sim 20$ kpc and effectively produces a plateau of constant dark matter density at the position of the BCG.

%%%%%%%%%%%%%%%%%%%%%%%%%%%%%%%%%%%%
\section{Conclusions}\label{sect_S8}
%%%%%%%%%%%%%%%%%%%%%%%%%%%%%%%%%%%%

Using the latest optical images of the cluster A370 from the HFF program (in the F435W, F606W and F814W filters) 
we have uncovered many new multiply lensed systems and constrained their redshifts geometrically, bringing the total number of 
(candidate) lensed background galaxies to 30 and the number of multiply lensed images to $\approx 80$.
We derive a set of mass models for the cluster, spanning a range of the most important variables, 
and compare the resulting mass profiles, projected mass distributions and critical curves. The models agree well with each other.
An NFW model with a small concentration parameter $C \approx 2$ agrees well with the observations (or a gNFW with $C \approx 4$  and inner slope $\gamma = 0.9$). A detailed analysis of the radial arcs near the BCGs points towards profiles with even smaller concentrations. We estimate the contribution to the mass density from the stellar component in the BCGs and find that in order to reproduce the observations, the inner slope of the dark matter density profile must be close to zero. We conclude that some mechanism must be acting in the two BCGs in order to expel most of the dark matter from them, or to avoid the formation of a cusp, or that dark matter is collisional.

%%%%%%%%%%%%%%%%%%%%%%%%%%%  
\section{Acknowledgments}  
%%%%%%%%%%%%%%%%%%%%%%%%%%%  
This work is based on observations made with the NASA/ESA {\it Hubble Space Telescope} and operated by the Association of Universities for Research in Astronomy, Inc. under NASA contract NAS 5-2655. 
Part of the data for this study is retrieved from the Mikulski Archive for Space Telescope (MAST).
The authors would like to thank the HFF team for making this spectacular data set promptly available to the community. Support for GLASS (HST-GO-13459) was provided by NASA through a grant from the Space Telescope Science Institute. 
The scientific results reported in this article are based in part on data obtained from the Chandra Data Archive \footnote{ivo://ADS/Sa.CXO\#obs/7715,ivo://ADS/Sa.CXO\#obs/555}. 
We thank the team from the MUSIC project for making available the simulations that have been used in this work. We would like to thank Harald Ebeling for making  the code {\small ASMOOTH} \citep{Ebeling2006} available. T. J. Broadhurst gratefully acknowledges the Visiting Research Professor Scheme at the University of Hong Kong. J. Lim acknowledges a grant from the Seed Funding for Basic Research of the University of Hong Kong (Project No. 201411159166) for this work.
J.M.D acknowledges support of the projects AYA2015-64508-P (MINECO/FEDER, UE), AYA2012-39475-C02-01 and the consolider project CSD2010-00064 funded by the Ministerio de Economia y Competitividad. 
%The work of JS was supported at IAP by  ERC Project No. 267117 (DARK) hosted by the Pierre and Marie Curie University- Paris 6, and at JHU by NSF grant OIA-1124403. 
JMD acknowledges the hospitality of the Physics Department at University of Pennsylvania for hosting him during the preparation of this work. The authors wish to give a special thank you to Tommaso Treu who carefully reviewed the paper and made important suggestions that helped us improved the contents of the paper. 

%%%%%%%%%%%%%%%%%%%%%%%%%%%%%%%%%%%%%%%%%%%%%%%%%%%%%%%%%%%%%%%%%%%%%%%  
  
\label{lastpage}
\bibliographystyle{mn2e}% style aa.bst 
\bibliography{MyBiblio} % References in MyBiblio.bib run bibtex after latex/pdflatex 

\appendix

\section{Compilation of arc positions}
This appendix presents the sample of secure and likely lensed multiple images detected behind A370 using the updated imaging
fron the HFF program, and spectroscopic redshifts from GLASS and the literature. 
Table~\ref{tab_arcs} lists the complete sample of images and their redshifts assigning IDs 
to each of them. 

The first column shows system ID following the original notation of \cite{Richard2014} and \cite{Johnson2014} 
(ID1.ID2.ID3 = System.Image.Knot) and ranks (A, B and C) . 
The redshifts $z_{\rm spect}$ are obtained from \cite{Richard2010,Richard2014,Johnson2014}.
GLASS redshifts are given in the column  $z_{\rm GLASS}$ while the photometric redshifts are given in column $z_{\rm BPZ}$. 
The systems having spectroscopic redshift are marked with bold face. 
Redshifts predicted by the lens model are given in column  $z_{\rm model}$. 
The column labeled Rank shows the quality of the system. Systems marked with rank (A) are very reliable and are used to derive the {\it driver} model. 
Systems marked with (B) are used to derive (together with systems having rank A) an alternative model. 
Systems with rank (B) are still reliable but their redshifts may be less precise than systems with rank (A). 
Systems marked with (C) are less reliable but still highly consistent with the driver lens model.  
In the last column 1, 2 and 3 refer to previous work where these systems are defined.  
1 stands for \citep{Richard2010}, 2 for \citep{Richard2014} and 3 for \citep{Johnson2014}
%%%%%%%%%%%%%%%%%%%%%%%%%%%%%%%%%%%%%%%%%%%%%%%%%%%%%%%%%%%%%%%%%%%%%%%%%%%%%%%%%%%%%%%%%%%%%%

\begin{table*}
  \begin{minipage}{165mm}                                               
    \caption{Full strong lensing data set. See text for description of the columns. 
             For the photometric redshifts we indicate the range of redshifts (from multiple images) after excluding extreme values.}
 \label{tab_arcs}
 \begin{tabular}{|ccccccccccc|}   
 \hline
%--------------------------------------------------------------------------------------------------------------------------------------------------
   KnotID   &       RA       &     DEC  & $z_{\rm used}$ & Old $z_{\rm spect}$ & $z_{\rm GLASS}$ & $z_{\rm BPZ}$ &  $z_{\rm EAZY}$ & $z_{\rm model}$ & Rank & Comments \\
%---------------------------------------------------------------------------------------------------------------------------------------------------
 \hline
{\bf 1.1.1}   &  2 39 54.310     & -1 34 33.75    &      0.806       &  0.806     &   0.8    & 0.805 -- 0.83  &   0.78 -- 0.8   &             &   (A)  &  1,2,3  \\
{\bf 1.2.1}   &  2 39 52.100     & -1 34 36.86    &                  &            &          &                &      &             &   (A)  &  1,2,3  \\
{\bf 1.3.1}   &  2 39 52.484     & -1 34 35.75    &                  &            &          &                &      &             &   (A)  &  1,2,3  \\
{\bf 1.1.2}   &  2 39 54.333     & -1 34 33.40    &                  &            &          &                &      &             &   (A)  &  1,2,3  \\
{\bf 1.2.2}   &  2 39 52.209     & -1 34 36.38    &                  &            &          &                &      &             &   (A)  &  1,2,3  \\
{\bf 1.3.2}   &  2 39 52.352     & -1 34 35.92    &                  &            &          &                &      &             &   (A)  &  1,2,3  \\
 \hline
{\bf 2.1.1}   &  2 39 53.724     & -1 35 03.21    &  0.725           &   0.725    &   0.71   & 0.58  -- 0.73  &    0.5  -- 1.0  &             &   (A)  &  1,2,3  \\
{\bf 2.2.1}   &  2 39 53.029     & -1 35 06.17    &                  &            &          &                &      &             &   (A)  &  1,2,3  \\
{\bf 2.3.1}   &  2 39 52.499     & -1 35 04.27    &                  &            &          &                &      &             &   (A)  &  1,2,3  \\
{\bf 2.4.1}   &  2 39 52.662     & -1 35 05.04    &                  &            &          &                &      &             &   (A)  &  1,2,3  \\
{\bf 2.5.1}   &  2 39 52.715     & -1 35 05.43    &                  &            &          &                &      &             &   (A)  &  1,2,3  \\
{\bf 2.1.2}   &  2 39 53.608     & -1 35 03.64    &                  &            &          &                &      &             &   (A)  &  1,2,3  \\
{\bf 2.1.3}   &  2 39 53.629     & -1 35 03.90    &                  &            &          &                &      &             &   (A)  &  1,2,3  \\
{\bf 2.1.4}   &  2 39 53.523     & -1 35 04.43    &                  &            &          &                &      &             &   (A)  &  1,2,3  \\
{\bf 2.1.5}   &  2 39 53.718     & -1 35 03.70    &                  &            &          &                &      &             &   (A)  &  1,2,3  \\
{\bf 2.1.6}   &  2 39 53.782     & -1 35 03.71    &                  &            &          &                &      &             &   (A)  &  1,2,3  \\
{\bf 2.1.7}   &  2 39 53.835     & -1 35 03.46    &                  &            &          &                &      &             &   (A)  &  1,2,3  \\
{\bf 2.1.8}   &  2 39 53.819     & -1 35 02.70    &                  &            &          &                &      &             &   (A)  &  1,2,3  \\
{\bf 2.1.9}   &  2 39 53.749     & -1 35 02.06    &                  &            &          &                &      &             &   (A)  &  1,2,3  \\
{\bf 2.1.10}  &  2 39 53.528     & -1 35 03.64    &                  &            &          &                &      &             &   (A)  &  1,2,3  \\
{\bf 2.2.2}   &  2 39 53.383     & -1 35 05.41    &                  &            &          &                &      &             &   (A)  &  1,2,3  \\
{\bf 2.2.3}   &  2 39 53.305     & -1 35 05.83    &                  &            &          &                &      &             &   (A)  &  1,2,3  \\
{\bf 2.2.4}   &  2 39 53.429     & -1 35 05.25    &                  &            &          &                &      &             &   (A)  &  1,2,3  \\
{\bf 2.3.2}   &  2 39 52.379     & -1 35 03.34    &                  &            &          &                &      &             &   (A)  &  1,2,3  \\
{\bf 2.3.3}   &  2 39 52.429     & -1 35 04.14    &                  &            &          &                &      &             &   (A)  &  1,2,3  \\
{\bf 2.3.4}   &  2 39 52.332     & -1 35 03.29    &                  &            &          &                &      &             &   (A)  &  1,2,3  \\
{\bf 2.3.9}   &  2 39 52.422     & -1 35 02.97    &                  &            &          &                &      &             &   (A)  &  1,2,3  \\
{\bf 2.3.10}  &  2 39 52.253     & -1 35 02.08    &                  &            &          &                &      &             &   (A)  &  1,2,3  \\
 \hline
{\bf 3.1.1}   &  2 39 54.543     & -1 34 01.89    &  1.95            &  1.42      &  1.95    &  1.4 -- 2.0    &    1.7 -- 1.9   &             &   (A)  &  1,2,3  \\
{\bf 3.2.1}   &  2 39 52.447     & -1 33 56.92    &                  &            &          &                &      &             &   (A)  &  1,2,3  \\
{\bf 3.3.1}   &  2 39 51.756     & -1 34 00.68    &                  &            &          &                &      &             &   (A)  &  1,2,3  \\
{\bf 3.1.2}   &  2 39 54.546     & -1 34 01.79    &                  &            &          &                &      &             &   (A)  &  1,2,3  \\
{\bf 3.2.2}   &  2 39 52.335     & -1 33 57.22    &                  &            &          &                &      &             &   (A)  &  1,2,3  \\
{\bf 3.3.2}   &  2 39 51.813     & -1 34 00.26    &                  &            &          &                &      &             &   (A)  &  1,2,3  \\
{\bf 3.1.3}   &  2 39 54.431     & -1 34 01.23    &                  &            &          &                &      &             &   (A)  &  1,2,3  \\
{\bf 3.2.3}   &  2 39 52.509     & -1 33 56.65    &                  &            &          &                &      &             &   (A)  &  1,2,3  \\
{\bf 3.3.3}   &  2 39 51.487     & -1 34 03.19    &                  &            &          &                &      &             &   (A)  &  1,2,3  \\
{\bf 3.1.4}   &  2 39 54.607     & -1 34 02.07    &                  &            &          &                &      &             &   (A)  &  1,2,3  \\
{\bf 3.2.4}   &  2 39 52.155     & -1 33 58.03    &                  &            &          &                &      &             &   (A)  &  1,2,3  \\
{\bf 3.3.4}   &  2 39 51.883     & -1 33 59.74    &                  &            &          &                &      &             &   (A)  &  1,2,3  \\
 \hline
{\bf 4.1.1}   &  2 39 55.116     & -1 34 34.99    &  1.27            &  1.27      &  1.272   & 1.13 -- 2.4    &   1.3 -- 1.7      &             &   (A)  &  1,2,3  \\
{\bf 4.2.1}   &  2 39 52.973     & -1 34 34.57    &                  &            &          &                &      &             &   (A)  &  1,2,3  \\
{\bf 4.3.1}   &  2 39 50.865     & -1 34 40.57    &                  &            &          &                &      &             &   (A)  &  1,2,3  \\
{\bf 4.1.2}   &  2 39 55.137     & -1 34 35.51    &                  &            &          &                &      &             &   (A)  &  1,2,3  \\
{\bf 4.1.3}   &  2 39 55.089     & -1 34 34.21    &                  &            &          &                &      &             &   (A)  &  1,2,3  \\
{\bf 4.2.2}   &  2 39 52.956     & -1 34 35.09    &                  &            &          &                &      &             &   (A)  &  1,2,3  \\
{\bf 4.2.3}   &  2 39 52.984     & -1 34 33.79    &                  &            &          &                &      &             &   (A)  &  1,2,3  \\
{\bf 4.3.2}   &  2 39 50.879     & -1 34 41.04    &                  &            &          &                &      &             &   (A)  &  1,2,3  \\
{\bf 4.3.3}   &  2 39 50.851     & -1 34 39.90    &                  &            &          &                &      &             &   (A)  &  1,2,3  \\
 \hline
     5.1.1    &  2 39 53.632     & -1 35 20.58    &  1.25            &            &          & 1.22 -- 1.53   &    1.3 -- 1.6   &  1.1        &   (B)  &  1,2,3  \\
     5.2.1    &  2 39 53.048     & -1 35 21.21    &                  &            &          &                &      &             &   (B)  &  1,2,3  \\
     5.3.1    &  2 39 52.563     & -1 35 20.61    &                  &            &          &                &      &             &   (B)  &  1,2,3  \\
     5.1.2    &  2 39 53.614     & -1 35 20.61    &                  &            &          &                &      &             &   (B)  &  1,2,3  \\
     5.1.3    &  2 39 53.658     & -1 35 20.49    &                  &            &          &                &      &             &   (B)  &  1,2,3  \\
     5.2.2    &  2 39 53.155     & -1 35 21.25    &                  &            &          &                &      &             &   (B)  &  1,2,3  \\
     5.2.3    &  2 39 52.928     & -1 35 21.10    &                  &            &          &                &      &             &   (B)  &  1,2,3  \\
     5.3.2    &  2 39 52.459     & -1 35 20.33    &                  &            &          &                &      &             &   (B)  &  1,2,3  \\
     5.3.3    &  2 39 52.711     & -1 35 20.82    &                  &            &          &                &      &             &   (B)  &  1,2,3  \\
 \end{tabular}  
 \end{minipage}
\end{table*}

\setcounter{table}{1}
 \begin{table*}
    \begin{minipage}{165mm}                                               
    \caption{cont.}
 \begin{tabular}{ccccccccccc}    
 \hline
%---------------------------------------------------------------------------------------------------------------------------------
   KnotID   &       RA       &     DEC  & $z_{\rm used}$ & Old $z_{\rm spect}$ & $z_{\rm GLASS}$ & $z_{\rm BPZ}$ &  $z_{\rm EAZY}$ & $z_{\rm model}$ & Rank & Comments \\
%---------------------------------------------------------------------------------------------------------------------------------
 \hline
{\bf 6.1.1}   &  2 39 52.662     & -1 34 37.94    &  1.063           &   1.063    &  1.32    & 1.0 -- 1.2    &    1.0 -- 1.2    &             &   (A)  &  1,2,3  \\
{\bf 6.2.1}   &  2 39 51.439     & -1 34 41.63    &                  &            &          &                &      &             &   (A)  &  1,2,3  \\
{\bf 6.3.1}   &  2 39 55.115     & -1 34 37.53    &                  &            &          &                &      &             &   (A)  &  1,2,3  \\
{\bf 6.2.2}   &  2 39 51.477     & -1 34 42.04    &                  &            &          &                &      &             &   (A)  &  1,2,3  \\
{\bf 6.2.3}   &  2 39 51.372     & -1 34 41.87    &                  &            &          &                &      &             &   (A)  &  1,2,3  \\
{\bf 6.1.2}   &  2 39 52.645     & -1 34 38.29    &                  &            &          &                &      &             &   (A)  &  1,2,3  \\
{\bf 6.1.3}   &  2 39 52.737     & -1 34 37.87    &                  &            &          &                &      &             &   (A)  &  1,2,3  \\
{\bf 6.3.2}   &  2 39 55.137     & -1 34 37.89    &                  &            &          &                &      &             &   (A)  &  1,2,3  \\
{\bf 6.3.3}   &  2 39 55.031     & -1 34 37.64    &                  &            &          &                &      &             &   (A)  &  1,2,3  \\
 \hline
     7.1.1    &  2 39 50.770     & -1 34 48.02    &  3.0	     &            &          & 2.9 -- 3.3     &    0.3 -- 0.8    &   2.8       &   (A)  &  1,2,3  \\
     7.2.1    &  2 39 52.769     & -1 34 50.78    &                  &            &          &                &      &             &   (A)  &  1,2,3  \\
     7.3.1    &  2 39 52.746     & -1 34 49.55    &                  &            &          &                &      &             &   (A)  &  1,2,3  \\
     7.4.1    &  2 39 52.514     & -1 35 08.28    &                  &            &          &                &      &             &   (A)  &  1,2,3  \\
     7.5.1    &  2 39 56.773     & -1 34 39.29    &                  &            &          &                &      &             &   (A)  &    \\
     7.6.1    &  2 39 52.569     & -1 34 38.92    &                  &            &          &                &      &             &   (C)  &    \\
     7.1.2    &  2 39 50.759     & -1 34 47.29    &                  &            &          &                &      &             &   (A)  &  1,2,3  \\
     7.2.2    &  2 39 52.815     & -1 34 53.33    &                  &            &          &                &      &             &   (A)  &  1,2,3  \\
     7.3.2    &  2 39 52.706     & -1 34 47.50    &                  &            &          &                &      &             &   (A)  &  1,2,3  \\
     7.4.2    &  2 39 52.544     & -1 35 08.19    &                  &            &          &                &      &             &   (A)  &  1,2,3  \\
     7.5.2    &  2 39 56.773     & -1 34 38.93    &                  &            &          &                &      &             &   (A)  &    \\
     7.6.2    &  2 39 52.569     & -1 34 37.93    &                  &            &          &                &      &             &   (C)  &    \\
 \hline
     8.1.1    &  2 39 51.473     & -1 34 11.37    &  3.0             &            &          & 2.9 -- 3.1     &  1.6 -- 3.0     &  2.0        &   (B)  &  1,2,3 \\
     8.2.1    &  2 39 50.856     & -1 34 25.10    &                  &            &          &                &      &             &   (B)  &  1,2,3 \\
     8.3.1    &  2 39 56.650     & -1 34 25.38    &                  &            &          &                &      &             &   (B)  &    \\
 \hline
{\bf 9.1.1}   &  2 39 50.976     & -1 34 40.39    &  1.52            &  1.52      &          & 1.6 -- 1.8     &  1.8 -- 1.9      &  1.55       &   (A)  &  1,2 \\
{\bf 9.2.1}   &  2 39 52.676     & -1 34 34.56    &                  &            &          &                &      &             &   (A)  &  1,2 \\
{\bf 9.3.1}   &  2 39 55.684     & -1 34 35.52    &                  &            &          &                &      &             &   (A)  &  1,2 \\
 \hline
     10.1.1   &  2 39 55.515     & -1 34 41.37    &  3.1             &            &          & 3.0 -- 3.2     &    2.4 -- 3.2    &  2.8        &   (B)  &    \\
     10.2.1   &  2 39 53.348     & -1 34 40.74    &                  &            &          &                &      &             &   (B)  &    \\
     10.3.1   &  2 39 53.804     & -1 35 08.19    &                  &            &          &                &      &             &   (B)  &    \\
     10.4.1   &  2 39 49.841     & -1 34 49.61    &                  &            &          &                &      &             &   (B)  &    \\
     10.1.2   &  2 39 55.508     & -1 34 41.68    &                  &            &          &                &      &             &   (B)  &    \\
     10.2.2   &  2 39 53.348     & -1 34 41.67    &                  &            &          &                &      &             &   (B)  &    \\
     10.3.2   &  2 39 53.832     & -1 35 08.09    &                  &            &          &                &      &             &   (B)  &    \\
 \hline
     11.1.1   &  2 39 51.313     & -1 34 09.70    &  5.9             &            &          &    --          &   --   &  $>$3       &   (B)  &  2  \\
     11.2.1   &  2 39 50.585     & -1 34 26.93    &                  &            &          &                &      &             &   (B)  &  2  \\
 \hline
     12.1.1   &  2 39 56.188     & -1 34 15.09    &  3.45            &            &          & 3.4 -- 3.5     &   3.3 -- 3.4   &  3.5        &   (B)  &  2  \\
     12.2.1   &  2 39 52.703     & -1 33 59.85    &                  &            &          &                &      &             &   (B)  &  2  \\
     12.3.1   &  2 39 50.213     & -1 34 30.88    &                  &            &          &                &      &             &   (B)  &  2  \\
 \hline
     13.1.1   &  2 39 55.089     & -1 34 18.44    &  4.0             &            &          & 3.97 --4.02    & 3.97 -- 4.02     &  $>$3       &   (B)  &  2  \\
     13.2.1   &  2 39 54.043     & -1 34 07.77    &                  &            &          &                &      &             &   (B)  &  2  \\
 \hline
     14.1.1   &  2 39 51.706     & -1 34 40.88    &    1.3           &            &          & 1.1 -- 1.6     &  0.9 -- 1.8    & 1.3         &   (A)  &    \\
     14.2.1   &  2 39 52.304     & -1 34 38.45    &                  &            &          &                &      &             &   (A)  &    \\
     14.3.1   &  2 39 55.751     & -1 34 37.11    &                  &            &          &                &      &             &   (A)  &    \\
 \hline
{\bf 15.1.1}  &  2 39 51.259     & -1 33 56.28    &  1.035           &            &   1.035  & 1.022          &     --    & 0.9         &   (A)  &    \\
{\bf 15.2.1}  &  2 39 51.110     & -1 33 58.00    &                  &            &          &                &      &             &   (A)  &    \\
{\bf 15.3.1}  &  2 39 51.210     & -1 33 57.43    &                  &            &          &                &      &             &   (A)  &    \\
 \hline
     16.1.1   &  2 39 50.868     & -1 34 59.41    &  3.75            &            &          & 3.7 -- 3.9     &   3.6 -- 3.9      &  $>$3       &   (A)  &    \\
     16.1.2   &  2 39 50.809     & -1 34 59.57    &                  &            &          &                &      &             &   (A)  &    \\
     16.1.3   &  2 39 50.816     & -1 34 58.37    &                  &            &          &                &      &             &   (A)  &    \\
     16.2.1   &  2 39 56.958     & -1 34 43.92    &                  &            &          &                &      &             &   (A)  &    \\
     16.2.2   &  2 39 56.900     & -1 34 44.77    &                  &            &          &                &      &             &   (A)  &    \\
     16.2.3   &  2 39 56.947     & -1 34 43.54    &                  &            &          &                &      &             &   (A)  &    \\
 \hline
     17.1.1   &  2 39 54.328     & -1 34 55.92    &  1.0             &            &          & 0.60 -- 0.65   &  0.868 --0.871    & 1.0         &   (C)  &    \\
     17.2.1   &  2 39 51.046     & -1 34 56.08    &                  &            &          &                &      &             &   (C)  &    \\
 \hline
     18.1.1   &  2 39 55.602     & -1 34 46.87    &  3.2             &            &          & 3.21 -- 3.23   &   2.2 -- 3.0    &  3.0        &   (B)  &    \\
     18.2.1   &  2 39 53.858     & -1 35 09.96    &                  &            &          &                &      &             &   (B)  &    \\
 \end{tabular}
 \end{minipage}
\end{table*}

\setcounter{table}{1}
 \begin{table*}
    \begin{minipage}{165mm}                                               
    \caption{cont.}
 \begin{tabular}{ccccccccccc}    
 \hline
%---------------------------------------------------------------------------------------------------------------------------------
   KnotID   &       RA       &     DEC  & $z_{\rm used}$ & Old $z_{\rm spect}$ & $z_{\rm GLASS}$ & $z_{\rm BPZ}$ &  $z_{\rm EAZY}$ & $z_{\rm model}$ & Rank & Comments \\
%---------------------------------------------------------------------------------------------------------------------------------
 \hline
     19.1.1   &  2 39 52.321     & -1 34 15.15    &  2.5             &            &          &     0.332      &  0.68    &  $>$2       &   (C)  &    \\
     19.2.1   &  2 39 52.456     & -1 34 18.34    &                  &            &          &                &      &             &   (C)  &    \\
 \hline
     20.1.1   &  2 39 51.147     & -1 34 14.14    &  4.7             &            &          & 4.71 -- 4.74   &   3.1 -- 4.2    &  $>$3       &   (C)  &    \\
     20.2.1   &  2 39 50.887     & -1 34 20.41    &                  &            &          &                &      &             &   (C)  &    \\
     20.1.2   &  2 39 51.137     & -1 34 14.52    &                  &            &          &                &      &             &   (C)  &    \\
     20.2.2   &  2 39 50.930     & -1 34 19.55    &                  &            &          &                &      &             &   (C)  &    \\
     20.1.3   &  2 39 51.182     & -1 34 13.29    &                  &            &          &                &      &             &   (C)  &    \\
     20.2.3   &  2 39 50.813     & -1 34 22.47    &                  &            &          &                &      &             &   (C)  &    \\
 \hline
     21.1.1   &  2 39 51.238     & -1 34 56.21    &  2.8             &            &          &  0.3 -- 3.2    &  0.4 -- 0.7    &  2.8        &   (C)  &    \\
     21.2.1   &  2 39 52.094     & -1 35 04.41    &                  &            &          &                &      &             &   (C)  &    \\
 \hline
     22.1.1   &  2 39 50.908     & -1 34 30.90    &  2.15            &            &          & 1.85 -- 2.15   & 1.81 -- 1.87     &  2.2        &   (B)  &    \\
     22.2.1   &  2 39 56.173     & -1 34 24.08    &                  &            &          &                &      &             &   (B)  &  8.3 in refs 1,2,3  \\
     22.3.1   &  2 39 51.975     & -1 34 10.81    &                  &            &          &                &      &             &   (B)  &    \\
 \hline
     23.1.1   &  2 39 50.676     & -1 34 30.91    &  2.3             &            &          & 0.275 -- 0.3   &  2.15 -- 2.25    &  2.3        &   (C)  &    \\
     23.2.1   &  2 39 56.110     & -1 34 22.42    &                  &            &          &                &      &             &   (C)  &    \\
     23.3.1   &  2 39 52.198     & -1 34 08.29    &                  &            &          &                &      &             &   (C)  &    \\
 \hline
     24.1.1   &  2 39 54.714     & -1 34 33.15    &  0.88            &            &  1.49    & 0.125 --0.495  &  0.42 -- 0.48    &  0.88       &   (C)  &    \\
     24.2.1   &  2 39 52.544     & -1 34 32.90    &                  &            &          &                &      &             &   (C)  &    \\
     24.3.1   &  2 39 51.901     & -1 34 35.45    &                  &            &          &                &      &             &   (C)  &    \\
 \hline
     25.1.1   &  2 39 54.652     & -1 34 58.05    &  2.45            &            &          & 2.43 -- 2.525  &  2.0    &  1.8        &   (C)  &    \\
     25.2.1   &  2 39 54.395     & -1 35 00.97    &                  &            &          &                &      &             &   (C)  &    \\
 \hline
     26.1.1   &  2 39 52.092     & -1 35 05.46    &  1.2             &            &          &     2.7        &  1.4    &  1.0 -- 1.4 &   (C)  &    \\
     26.2.1   &  2 39 52.172     & -1 35 06.10    &                  &            &          &                &      &             &   (C)  &    \\
     26.3.1   &  2 39 51.978     & -1 35 04.61    &                  &            &          &                &      &             &   (C)  &    \\
 \hline
     27.1.1   &  2 39 53.386     & -1 34 01.78    &  2.9             &            &          &  1.2 -- 2.9    &   0.3 -- 1.3   &   3.1       &   (B)  &    \\
     27.2.1   &  2 39 55.366     & -1 34 16.02    &                  &            &          &                &      &             &   (B)  &    \\
     27.3.1   &  2 39 50.254     & -1 34 23.94    &                  &            &          &                &      &             &   (B)  &    \\
 \hline
     28.1.1   &  2 39 52.344     & -1 33 53.37    &  4.5             &            &          &   4.5 -- 5.6   &  3.4 -- 4.4    &   $>$3      &   (C)  &    \\
     28.2.1   &  2 39 55.261     & -1 34 00.35    &                  &            &          &                &      &             &   (C)  &    \\
     28.3.1   &  2 39 50.618     & -1 34 08.47    &                  &            &          &                &      &             &   (C)  &    \\
 \hline
     29.1.1   &  2 39 51.665     & -1 35 16.09    &  5.5             &            &          &        --      &      --        &   $>$5     &    (C)  &    \\
     29.2.1   &  2 39 51.266     & -1 35 12.78    &                  &            &          &                &      &             &   (C)  &    \\
 \hline
     30.1.1   &  2 39 54.517     & -1 34 25.65    &  2.3             &            &          &        --      &      --        &   2.3     &    (C)  &    \\
     30.2.1   &  2 39 54.015     & -1 34 25.65    &                  &            &          &                &      &             &   (C)  &    \\

\end{tabular}  
 \end{minipage}
\end{table*}

\end{document}